\documentclass[acmsmall,nonacm]{acmart} 

\usepackage{array}
\usepackage{cleveref}
\usepackage{colortbl}
\usepackage{csquotes}
\usepackage{mdframed}
\usepackage{xspace}

\acmJournal{CSUR}

\hypersetup{
  pdftitle = {Anomaly Detection and Root Cause Analysis - Survey},
  pdfauthor = {Soldani J. and Brogi A.}
}

\newcommand{\eg}{e.g.,\xspace}
\newcommand{\etal}{\textit{et al.}\xspace}

\newcommand{\viz}{viz.,\xspace}

\newcommand{\upd}[1]{#1} 

\newcommand{\thickline}{\specialrule{1pt}{0.2pt}{0.2pt}}
\newcommand{\thinline}{\specialrule{.4pt}{0.1pt}{0.1pt}}

\begin{document}

\title{Anomaly Detection and Failure Root Cause Analysis in (Micro)Service-Based Cloud Applications: A Survey}

\renewcommand{\shorttitle}{Anomaly Detection and Failure Root Cause Analysis in (Micro)Service-Based Cloud Applications}

\author{Jacopo Soldani}
\email{jacopo.soldani@unipi.it}
\author{Antonio Brogi}
\email{antonio.brogi@unipi.it}
\affiliation{%
  \institution{University of Pisa}
  \streetaddress{Dipartimento di Informatica, Largo B. Pontecorvo 3}
  \city{Pisa}
  \state{PI}
  \country{Italy}
  \postcode{56127}
}

\renewcommand{\shortauthors}{J. Soldani and A. Brogi}

\begin{abstract}
The momentum gained by microservices and cloud-native software architecture pushed nowadays enterprise IT towards multi-service applications.
The proliferation of services and service interactions within applications, 
often consisting of hundreds of interacting services, makes it harder to detect failures and to identify their
possible root causes, which is on the other hand crucial to promptly recover and fix applications. Various techniques
have been proposed to promptly detect failures based on their symptoms, viz., observing anomalous
behaviour in one or more application services, as well as to analyse logs or monitored performance of such
services to determine the possible root causes for observed anomalies. 
The objective of this survey is to provide a structured overview and a qualitative analysis of currently available techniques for anomaly detection and root cause analysis in modern multi-service applications.
Some open challenges and research directions stemming out from the analysis are also discussed.
\end{abstract}




\maketitle

\section{Introduction}
\label{sec:intro}
With the rise of microservices, multi-service applications became the de-facto standard for delivering enterprise IT applications \cite{Soldani2018_MicroservicesPainsGains}.
Many big players are already delivering their core business through multi-service applications, with Amazon, Netflix, Spotify, and Twitter perhaps being the most prominent examples.
Multi-service applications, like microservice-based applications, hold the promise of exploiting the potentials of cloud computing to obtain cloud-native applications, \viz composed by loosely coupled services that can be indepedently deployed and scaled \cite{Kratzke2017_CloudNativeApplications}.

At the same time, services and service interactions proliferate in modern multi-service applications, often consisting of hundreds of interacting services.
This makes it harder to monitor the various services forming an application to detect whether they have failed, as well as to understand whether a service failed on its own or in cascade, \viz because some of the services it interacts with failed, causing the service to fail as well.
The detection and understanding of failures in modern multi-service applications is actually considered a concrete \enquote{pain} by their operators \cite{Soldani2018_MicroservicesPainsGains}.

Various solutions have been proposed to automate the detection of failures and to automatically determine their possible root causes.
Existing solutions for failure detection rely on identifying anomalies in the behaviour of services, which can be symptoms of their possible failures \cite{38_Nandi2016_AnomalyDetectionControlFlowGraphMining,34_Nedelkoski2019_AnomalyDetectionDistributedTracing,09_Wu2020_MicroserviceDiagnosisDeepLearning}.
Once an anomaly has been detected in a multi-service application, further analyses are enacted to determine the possible root causes for such an anomaly \cite{06_Lin2018_Microscope,49_Shan2019_epsilonDiagnosis}.
This allows application operators to determine whether the anomaly on a service was due to the service itself, to other services underperforming or failing as well, or to environmental reasons, \eg unforeseen peaks in user requests \cite{42_Nguyen2011_PAL} or lack of computing resources in the runtime environment \cite{40_Chen2020_CauseInfer}.

Existing solutions for anomaly detection and root cause analysis are however scattered across different pieces of literature, and often focus only on either anomaly detection or root cause analysis.
This hampers the work of application operators wishing to equip their multi-service applications with a pipeline for detecting anomalies and identifying their root causes.
To this end, in this article we survey the existing techniques for detecting anomalies in modern multi-service applications and for identifying the possible root causes of detected anomalies.
To further support an application operator in choosing the techniques most suited for her application, we also discuss the instrumentation needed to apply an anomaly detection/root cause analysis technique, as well as the additional artifacts that must be provided as input to such techniques.
We also highlight whether the surveyed techniques already put anomaly detection and root cause analysis in a pipeline, or whether they need to be integrated one another.
\upd{In the latter case, we comment on whether the type of anomalies that can be observed with a given anomaly detection technique can be explained by a given root cause analysis technique.}

We believe that our survey can provide benefits to both practitioners and researchers working with modern multi-service applications.
We indeed not only than help them in finding the anomaly detection and root cause analysis techniques most suited to their needs, but we also discuss some open challenges and possible research directions on the topic. 

\smallskip \noindent
The rest of this article is organised as follows.
\Cref{sec:terminology} introduces the terminology used in our survey.
\Cref{sec:detection,sec:rca} provide a structured overview of the existing techniques for detecting anomalies in multi-service applications and for identifying their root causes, respectively.
\Cref{sec:related} positions our survey with respect to related work.
\Cref{sec:conclusions} concludes the article by discussing possible research directions on detecting and explaining anomalies in multi-service applications.

\section{Terminology}
\label{sec:terminology}
Whilst \textit{failures} denote the actual inability of a service to perform its functions \cite{Saltzer2009_PrinciplesComputerSystemDesign}, \textit{anomalies} correspond to the observable symptoms for such failures \cite{29_Wang2018_CloudRanger}, \eg a service slowing its response time, reducing its throughput, or logging error events.   
The problem of \textit{anomaly detection} in multi-service applications hence consists of identifying anomalies that can possibly correspond to failures affecting their services.
Anomalies can be detected either at \textit{application-level} or at \textit{service-level}, based on whether symptoms of a failure are observed by considering the application as whole (\eg performance degradations or errors shown by its frontend) or by focusing on specific services.  

\textit{Root cause analysis} then consists of identifying the reasons why an application- or service-level anomaly has been observed, with the ultimate goal of providing possible reasons for the corresponding failure to occur.
Root cause analysis is enacted with techniques analysing what has actually happened while a multi-service application was running.
\upd{
    It is hence not to be confused with debugging techniques, which look for the possible reasons for an observed anomaly by re-running applications in a testing environment and by trying to replicate the observed anomaly (\eg as in the delta debugging proposed by Zhou \etal \cite{03_Zhou2019_DeltaDebugging}).
    Root cause analysis techniques instead only analyse information collected while the application was running in production (\eg logs or monitoring data), without re-running the application itself. 
}

Anomaly detection and root cause analysis are typically enacted based on runtime information collected on the services forming an application, hereafter also called \textit{application services}. 
Such information includes \textit{KPIs} (Key Performance Indicators) monitored on application services, \eg response time, availability, or resource consumption, as well as the \textit{events} logged by application services.
Each logged event provide information on something happened to the service logging such event, \eg whether it was experiencing some error, hence logging an \textit{error event}, or whether it was invoking or being invoked by another service.
The events logged by a service constitute the \textit{service logs}, whereas the union of all \textit{service logs} distributed among the services forming an application is hereafter referred as \textit{application logs}.

Application logs are not to be confused with \textit{distributed traces}, obtained by instrumenting application services to feature \textit{distributed tracing}.
The latter is a distributed logging method used to profile and monitor multi-service applications, which essentially consists of (i) instrumenting the services forming an application to assign each user request a unique id, (ii) passing such id to all services that are involved in processing the request, (iii) including the request id in all log messages, and (iv) recording information (\eg start time, end time) about the service invocations and internal operations performed when handling a request in a service \cite{Richardson2018_MicroservicesPatterns}.
All the events generated by user requests are hence tracked in the distributed \enquote{traces}, and traces of events generated by the same request can be identified based on the request id. 

\section{Anomaly Detection}
\label{sec:detection}
We hereafter survey the existing techniques for detecting anomalies in multi-service applications, \viz symptoms of failures affecting the whole application or some of its services.
We organise the survey by distinguishing three main classes of techniques:
we separate the techniques enacting anomaly detection by directly processing the logs produced by application services (\Cref{sec:detection:service-logs}) from those requiring to instrument applications, \viz to feature distributed tracing (\Cref{sec:detection:distributed-tracing}) or by installing agents to monitor application services (\Cref{sec:detection:monitoring}).
We finally discuss the surveyed anomaly detection techniques, with the support of a recapping table (\Cref{tab:detection}) and by also highlighting some open challenges in anomaly detection (\Cref{sec:detection:discussion}).

\subsection{Log-based Anomaly Detection Techniques}
\label{sec:detection:service-logs}

Log-based online anomaly detection is typically enacted by exploiting unsupervised machine learning algorithms, which are used to process the logs produced by the services forming an application to detect whether some anomaly is occurring in some application service.

\subsubsection{Unsupervised Learning}
Online anomaly detection in application services can be enacted by exploiting unsupervised learning algorithms to learn a baseline modelling of the logging behaviour of the application services in failure-free runs of an application.
The baseline model is then used online to detect whether the events newly logged by a service diverge from the baseline model, hence indicating that the service suffers from some anomaly.
This technique is adopted by OASIS \cite{38_Nandi2016_AnomalyDetectionControlFlowGraphMining}, Jia \etal \cite{31_Jia2017_AnomalyDiagnosisHybridGraphModel}, and LogSed \etal \cite{32_Jia2017_LogSed}, which we discuss hereafter.

OASIS \cite{38_Nandi2016_AnomalyDetectionControlFlowGraphMining} starts from the application logs generated in failure-free runs of the application, which are used as the training data to mine a control-flow graph modelling the baseline application behaviour under normal conditions, \viz which events should be logged by which service, and in which order.
OASIS \cite{38_Nandi2016_AnomalyDetectionControlFlowGraphMining} mines log templates from the textual content in the training logs, and it clusters the log templates so that all templates originated by the same log statement go in the same cluster.
The nodes in the control flow graph are then determined by picking a representative template from each cluster.
The edges are instead mined based on temporal co-occurrence of log templates, upon which OASIS \cite{38_Nandi2016_AnomalyDetectionControlFlowGraphMining} determines the immediate successors of each template, together with branching probabilities and expected time lags between each template and its successors.
Online anomaly detection is then enacted by mapping newly generated logs to their corresponding templates and by checking whether the arriving log templates adhere to the control-flow graph.
If none of the templates that should follow a template is logged in the corresponding time lag, or if the actual rate of templates following a given template significantly deviates from the expected branching probabilities, OASIS \cite{38_Nandi2016_AnomalyDetectionControlFlowGraphMining} considers the application as affected by a functional anomaly.

Jia \etal \cite{31_Jia2017_AnomalyDiagnosisHybridGraphModel} and LogSed \etal \cite{32_Jia2017_LogSed} provide two other techniques for the detection of functional or performance anomalies in multi-service applications based on the analysis of their logs.
They process the logs collected in normal, failure-free runs of an application to mine, for each service, a time-weighted control flow graph modelling its internal flow.
In particular, Jia \etal \cite{31_Jia2017_AnomalyDiagnosisHybridGraphModel} and LogSed \etal \cite{32_Jia2017_LogSed} learn the time-weighted control flow graph for each service, with a similar method to that in OASIS \cite{38_Nandi2016_AnomalyDetectionControlFlowGraphMining}:
template mining is used to transform logs into templates, which are then clustered based on events they correspond to, and the log clusters are processed to infer the sequencing relationships among events. 
The resulting graph of events is then enriched by weighting each arc with the average time passing between the logging of the source event and that of the target event.
The topology graph and the time-weighted control flow graphs constitute the baseline model against which to compare newly logged events. 
This is done by mapping the events logged by each service to their corresponding template and by checking whether its actual behaviour deviates from the expected one.
A service suffers from functional anomalies whenever it does not log an event that it should have been logging, or when unexpected logs occur.
A service instead suffers from performance anomalies when it logs an expected event significantly later than when expected. 

\subsection{Distributed Tracing-based Anomaly Detection Techniques}
\label{sec:detection:distributed-tracing}
Online anomaly detection is enacted also by instrumenting the target applications to feature additional functionalities, with the most common being distributed tracing.
Distributed tracing is indeed at the basis of various techniques for anomaly detection, combined with supervised or unsupervised machine learning,\footnote{Whilst unsupervised learning algorithms learn patterns directly from data, supervised learning algorithms rely on data to be labelled, \viz training examples associating given input data with desired output values \cite{Mitchell1997_MachineLearning}.} or with trace comparison techniques.

\subsubsection{Unsupervised Learning}
\label{sec:detection:distributed-tracing:unsupervised-learning}
TraceAnomaly \cite{16_Liu2020_TraceAnomaly} and Nedelkoski \etal \cite{35_Nedelkoski2019_AnomalyDetectionTracingData} collect traces in training runs of a multi-service application and use them to train unsupervised neural networks.
They then enact online anomaly detection by exploiting the trained neural networks to process the traces generated by the application while it is running in production.
In particular, TraceAnomaly \cite{16_Liu2020_TraceAnomaly} trains a deep Bayesian neural network with posterior flow \cite{Neal1996_BayesianNetwork,Rezende2015_NormalizingFlows}, which enables associating monitored traces with a likelihood to be normal, \viz not affected by performance anomalies.
It also stores all seen service call paths, \viz all sequences of service interactions observed in the available traces. 
TraceAnomaly \cite{16_Liu2020_TraceAnomaly} then enacts online anomaly detection in two steps.
It first checks whether a newly produced trace contains previously unseen call paths.
If this is the case, TraceAnomaly \cite{16_Liu2020_TraceAnomaly} decides whether to consider the unseen call paths as functional anomalies for the application, based on whether they appear in a whitelist manually filled by the application operator to exclude false alarms (\eg unseen call paths corresponding to new interactions occurring after service upgrades).
If there is no functional anomaly, the trace is passed to the deep Bayesian neural network, which associates the monitored trace with its likelihood to be normal.
If the likelihood is below a given threshold, the trace denotes a performance anomaly for the application.

Nedelkoski \etal \cite{35_Nedelkoski2019_AnomalyDetectionTracingData} train a multi-modal long-short term memory neural network \cite{Goodfellow2016_DeepLearningBook} modelling the normal behaviour of a multi-service application, based on the traces collected in failure-free runs of the application.
The long-short term memory neural network is multi-modal in the sense that it considers both the types of events logged in a trace and the response times of application services.
It is obtained by combining two single-modal networks, one trained to predict the probability distribution of the events following an event in a trace, and the other trained to predict the probability distribution of the response times for a service invocation.
The trained network is then used in online anomaly detection to predict the most probable events following an event and the most probable response times for a service invocation. 
If the event following another, or if the response time of a service are not amongst the predicted ones, the application is considered to suffer from a functional or performance anomaly, respectively.

Jin \etal \cite{22_Jin2020_AnomalyDetectionMicroservicesRPCA} allows detecting performance anomalies in multi-service applications based on the offline analysis of the traces collected through distributed tracing, by also considering the performance metrics of application services that can be directly collected from their runtime environment (\eg CPU and memory consumption).
Jin \etal \cite{22_Jin2020_AnomalyDetectionMicroservicesRPCA} first enact principal component analysis on logged traces to determine the services that may be involved in anomalous interactions.
This is done by deriving a matrix-based representation of the logged traces, by reducing the obtained matrix to its principal components, and by applying a function linearly combining such principal components to elicit the services involved in anomalous traces. 
The performance metrics collected for such services are then processed with unsupervised learning algorithms for anomaly detection, \viz isolation forest, one-class support vector machine, local outlier factor, or 3$\sigma$ \cite{Chandola2009_SurveyAnomalyDetection}. 
Such algorithms are used to determine anomalous values for such metrics, which are used to assign an anomaly score to the services involved in anomalous traces.
Jin \etal \cite{22_Jin2020_AnomalyDetectionMicroservicesRPCA} then return the list of services affected by performance anomalies, \viz the services whose anomaly score is higher than a given threshold, ranked by anomaly score.

\subsubsection{Supervised Learning}
Seer \cite{17_Gan2019_Seer}, Nedelkoski \etal \cite{34_Nedelkoski2019_AnomalyDetectionDistributedTracing}, and Bogatinovski \etal \cite{36_Bogatinovski2020_SelfSupervisedAnomalyDetection} also enact a combination of distributed tracing and deep learning, however following a supervised learning approach.
They indeed collect traces in training runs of a multi-service application, labelled to distinguish normal runs from those where anomalies are known to have occurred on specific services.
The collected traces are used to train neural networks, which are then used to enact online anomaly detection by processing the traces newly generated by the application.
In particular, Seer \cite{17_Gan2019_Seer} trains a deep neural network consisting of convolutional layers followed by long-short term memory layers \cite{Goodfellow2016_DeepLearningBook}. 
Input and output neurons of the network corresponds to application services.
Input neurons are used to pass the KPIs of each service, \viz latency and outstanding requests, logged by the service itself in distributed traces, and resource consumption, obtained by interacting with the nodes where the services are running. 
Output neurons instead indicate which services are affected by performance anomalies.
At runtime, Seer \cite{17_Gan2019_Seer} continuously feeds the trained deep neural network with the stream of monitored traces, and the network consumes such traces to determine whether some application service is affected by a performance anomaly. 
If this is the case, Seer \cite{17_Gan2019_Seer} interacts with the runtime of the node hosting the anomalous service to determine which computing resources are being saturated.
Seer \cite{17_Gan2019_Seer} then notifies the system manager about the resource saturation happening on the node, to allow the manager to mitigate the effects of the detected performance degradation, \eg by providing the node with additional computing resources.

Nedelkoski \etal \cite{34_Nedelkoski2019_AnomalyDetectionDistributedTracing} train two neural networks, \viz a variational autoencoder \cite{Kingma2015_VariationalAutoencoder} modelling the normal behaviour of the application and a convolutional neural network \cite{Goodfellow2016_DeepLearningBook} recognising the type of failure affecting a service when the latter is affected by a performance anomaly. 
The variational autoencoder is learned from the distributed traces obtained in normal runs of the application, so that the autoencoder learns to encode non-anomalous traces and to suitably reconstruct them from their encoding.
If applied on an anomalous trace, instead, the autoencoder reconstruct them with significant errors, hence enabling to detect anomalies.
The convolutional neural network is instead trained on the traces collected in runs of the application where specific failures were injected in specific services, so that the network can recognise which failures caused a performance anomaly in a service.
In online anomaly detection, the trained networks are used in a pipeline.
The outputs of the autoencoder are passed to a post-processor to exclude false positives, \viz performance anomalies affecting a service due to temporary reasons (\eg overload of a service). 
If a service is considered anomalous by the post-processor as well, the anomalous trace is passed to the convolutional network to detect the type of anomaly affecting the service. 

Bogatinovski \etal \cite{36_Bogatinovski2020_SelfSupervisedAnomalyDetection} work under a different assumption if compared with the other anomaly detection techniques discussed above, \viz it assumes that the logging of an event in a particular position in a trace is conditioned both by those logged earlier and by those appearing afterwards.
Bogatinovski \etal \cite{36_Bogatinovski2020_SelfSupervisedAnomalyDetection} indeed train a self-supervised encoder-decoder neural network \cite{Goodfellow2016_DeepLearningBook} capable of predicting the logging of an event in a given \enquote{masked} position of a trace based on the context given by its neighbour events. 
In particular, the trained neural network provides a probability distribution of the possible events appearing in any masked position of an input trace, given the context of the events appearing before and after the masked event to be predicted. 
The trained neural network is then used to enact anomaly detection over novel traces, \viz to obtain a sorted list of events that should have been logged in each position of the trace. 
The lists generated by the network are then analysed by a post-processor, which considers a truly logged event as anomalous if it is not amongst the events predicted to appear in the corresponding position. 
The post-processor then computes an anomaly score for the trace (\viz number of anomalous events divided by trace length).
If the anomaly score is beyond an user-specified threshold, the trace is considered to witness a functional anomaly affecting the application.

Finally, MEPFL \cite{02_Zhou2019_LatentErrorPrediction} provides another technique applying supervised machine learning to perform online anomaly detection on application instrumented to feature distributed tracing.
MEPFL \cite{02_Zhou2019_LatentErrorPrediction} can recognize functional anomalies by exploiting multiple supervised learning algorithms, \viz k-nearest neighbors \cite{Altman1992_NearestNeighbor}, random forests \cite{Breiman2001_RandomForests}, and multi-layer perceptron \cite{Goodfellow2016_DeepLearningBook}.
MEPFL \cite{02_Zhou2019_LatentErrorPrediction} takes as input a multi-service application and a set of automated test cases simulating user requests to load the application and to produce traces.
The traces must include information on the configuration of a service (\viz memory and CPU limits, and volume availability), as well as on the status, resource consumption, and service interactions of its instances.
This information is fed to one of the supported supervised learning algorithms, which trains classifiers determining whether a trace includes anomalies, which services are affected by such anomalies, and which types of failures caused such services to experience anomalies. 
The training of the classifiers is based on traces collected in training runs of an application, where the application is loaded with the input test cases, and where the application is run both in normal conditions and by injecting failures in its services to observe how the traces change when such type of failures affect such services.

\subsubsection{Trace Comparison}
Trace comparison is another possible technique to enact online anomaly detection on multi-service applications instrumented to feature distributed tracing.
This is the technique followed by Meng \etal \cite{23_Meng2021_DetectingAnomaliesMicroservices}, Wang \etal \cite{26_Wang2020_WorkflowAwareFaultDiagnosis}, and Chen \etal \cite{39_Chen2020_MatrixSketchBasedAnomalyDetection}, which all rely on collecting the traces that can possibly occur in a multi-service application and by then checking whether newly collected traces are similar to the collected ones.

Meng \etal \cite{23_Meng2021_DetectingAnomaliesMicroservices} and Wang \etal \cite{26_Wang2020_WorkflowAwareFaultDiagnosis} enable detecting functional anomalies by collecting traces while testing an application in a pre-production environment and by building a set of representative call trees from collected traces.
Such trees model the service invocation chains that can possibly appear in normal conditions, as they have been observed in the collected traces. 
Newly monitored traces are then reduced to their call trees, whose tree-edit distance from each representative call tree is computed with the RTDM algorithm \cite{Reis2004_RTDM}.
If the minimum among all the computed distances is higher than a given threshold, the trace is considered as anomalous, and the service from which tree edits start is identified as the service suffering from a functional anomaly.
Meng \etal \cite{23_Meng2021_DetectingAnomaliesMicroservices} and Wang \etal \cite{26_Wang2020_WorkflowAwareFaultDiagnosis} also enable detecting performance anomalies based on the services' response time.
The response times in the traces collected in the testing phase are modelled as a matrix, whose elements represent the response time for a service in each of the collected traces.
Principal component analysis \cite{Jolliffe2011_PCA} is then enacted to reduce the matrix to its principal components, which are then used to define a function determining whether the response time of a service in newly monitored traces is anomalous.
At the same time, as both Meng \etal \cite{23_Meng2021_DetectingAnomaliesMicroservices} and Wang \etal \cite{20_Wang2020_RootCauseMetricLocation} explicitly notice, the computational complexity of the enacted trace comparisons makes them better suited to enact offline anomaly detection, as they would be too time consuming to perform online anomaly detection on medium-/large-scale multi-service applications. 

Chen \etal \cite{39_Chen2020_MatrixSketchBasedAnomalyDetection} instead uses matrix sketching to compare traces and detect anomalies.
More precisely, Chen \etal \cite{39_Chen2020_MatrixSketchBasedAnomalyDetection} adapt an existing matrix sketching algorithm \cite{Huang2015_MatrixSketching} to detect response time anomalies in services in monitored traces. 
This is done by maintaining up-to-date a \enquote{sketch} of the high-dimensional data corresponding to the historically monitored response times of the services forming an application.
The sketch is a limited set of orthogonal basis vectors that can linearly reconstruct the space containing the non-anomalous response times for each service in the formerly monitored traces. 
The response time for a service in a newly monitored trace is non-anomalous if it lies within or close enough to the space defined by the sketch vector, given a tolerance threshold.
If this is the case, the sketch vectors are updated.
Otherwise, the corresponding service is considered to experience a performance anomaly.

\subsection{Monitoring-based Anomaly Detection Techniques}
\label{sec:detection:monitoring}
Online anomaly detection in multi-service applications can also be enacted by installing agents to monitor KPIs on their services, and by then processing such KPIs to detect anomalies.
In this perspective, a basic solution is comparing the KPIs monitored on the service acting as the application frontend against the application's SLOs: in case of SLO violations, the whole application is considered to suffer from a performance anomaly.
For detecting anomalies at a finer granularity, \viz going from application-level anomalies to service-level anomalies, the most common technique is to exploit unsupervised or supervised machine learning algorithms to process monitored KPIs and detect whether any anomaly is occurring in any application service.
Another possibility is to exploit self-adaptive heartbeat protocols to detect anomalous services, rather than processing their KPIs. 
We hereafter discuss all such techniques by starting from those based on unsupervised/supervised learning, which are then followed by those based on SLO checks and heartbeating.

\subsubsection{Unsupervised Learning}
\label{sec:detection:monitoring:unsupervised-learning}
Online anomaly detection can be enacted by exploiting unsupervised learning algorithms to process the KPIs monitored on application services in failure-free runs of the application and learn a baseline modelling of their behaviour.
The baseline model is then used online to detect whether the newly monitored KPIs on a service diverge from the baseline model, hence indicating that the service suffers from some anomaly.
This technique is adopted by Gulenko \etal \cite{37_Gulenko2018_DetectingAnomalousBehaviourBlackBoxServices}, MicroRCA \cite{04_Wu2020_MicroRCA}, Wu \etal \cite{09_Wu2020_MicroserviceDiagnosisDeepLearning}, LOUD \cite{13_Mariani2018_LocalizingFaultsCloudSystems}, and DLA \cite{14_Samir2019_DLA}, which train the baseline model by considering the KPIs monitored in normal runs of the application, \viz assuming that no anomaly occurred in such runs.

Gulenko \etal \cite{37_Gulenko2018_DetectingAnomalousBehaviourBlackBoxServices} and LOUD \cite{13_Mariani2018_LocalizingFaultsCloudSystems} are \enquote{KPI-agnostic}, meaning that they can work on any set of KPIs for the services forming an application, monitored with probes installed in the hosts where the application services are run.  
In particular, Gulenko \etal \cite{37_Gulenko2018_DetectingAnomalousBehaviourBlackBoxServices} enable detecting performance anomalies in multi-service applications by building different baseline models for different services.
The KPIs monitored on a service at a given time are modelled as a vector, and the BIRCH \cite{Zhang1996_BIRCH} online clustering algorithm is used in an initial training phase to cluster the vectors corresponding to the monitored KPIs.
At the end of the training phase, the centroids and centroid radii of the obtained clusters determine the baseline subspaces where the vectors of monitored KPIs should pertain to be classified as \enquote{normal}. 
This enables the online processing of newly monitored KPIs for each service, by simply checking whether the corresponding vector falls within any of the baseline subspaces, \viz whether there exists a cluster whose centroid is less distant from the vector of monitored KPIs than the corresponding centroid radius. 
If this is not the case, the corresponding service is considered as affected by a performance anomaly.

LOUD \cite{13_Mariani2018_LocalizingFaultsCloudSystems} instead trains a baseline model for the whole application by pairing each monitored KPI with the corresponding service.
The KPIs monitored in the offline runs of the application are passed to the IBM ITOA-PI \cite{IBM_ITOA_PI} to build a KPI baseline model and a causality graph.
The baseline model provides information about the average values and the acceptable variations over time for a KPI.
Each node in the causality graph correspond to a service's KPI, and the edges indicate the causal relationships among KPIs, with weights indicating the probability to which changes in the source KPI can cause changes in the target KPI (computed with the Granger causality test \cite{Arnold2007_GrangerMethodsTemporalCausalModelling}).
During the online phase, LOUD \cite{13_Mariani2018_LocalizingFaultsCloudSystems} again exploits the IBM ITOA-PI to detect performance anomalies: a KPI and the corresponding service are reported as anomalous if the monitored value for such KPI is outside of the acceptable variations coded in the baseline model, or if the causal relationships represented in the causality graph are not respected (\eg when a KPI significantly changed its value, but those that should have changed as well remained unchanged).

Differently from the above discussed, \enquote{KPI-agnostic} techniques \cite{37_Gulenko2018_DetectingAnomalousBehaviourBlackBoxServices,13_Mariani2018_LocalizingFaultsCloudSystems}, MicroRCA \cite{04_Wu2020_MicroRCA}, Wu \etal \cite{09_Wu2020_MicroserviceDiagnosisDeepLearning}, and DLA \cite{14_Samir2019_DLA} focus on given KPIs.
MicroRCA \cite{04_Wu2020_MicroRCA} and Wu \etal \cite{09_Wu2020_MicroserviceDiagnosisDeepLearning} consider the response time and resource consumption of application services, which they monitor by requiring the Kubernetes (k8s) deployment of multi-service applications to be instrumented as service meshes featuring Istio \cite{Istio} and Prometheus \cite{Prometheus}.
The latter are used to collect KPIs from the application services. 
MicroRCA \cite{04_Wu2020_MicroRCA} and Wu \etal \cite{09_Wu2020_MicroserviceDiagnosisDeepLearning} then follow an unsupervised learning approach similar to that proposed by Gulenko \etal \cite{37_Gulenko2018_DetectingAnomalousBehaviourBlackBoxServices}.
The KPIs monitored on each service 
are modelled as a vector, and the BIRCH algorithm \cite{Zhang1996_BIRCH} 
is used to cluster the vectors corresponding to KPIs monitored in normal runs of the application, \viz assuming that no anomaly occurred in such runs.
The obtained clusters determine the baseline subspaces where the vectors of monitored KPIs should pertain to be classified as \enquote{normal}. 
This enables the online processing of newly monitored KPIs for each service, by simply checking whether the corresponding vector falls within any of the baseline subspaces. 
If this is not the case, the service is considered as affected by a performance anomaly.

DLA \cite{14_Samir2019_DLA} instead focuses on the response time of each application service, whilst also considering its possible fluctuations due to increase/decrease of end users' transactions.
It also focuses on containerised multi-service applications, whose deployment in k8s is expected to be provided by the application operator. 
DLA \cite{14_Samir2019_DLA} installs a monitoring agent in the VMs used to deploy the application.
The agents collect the response time of the containerised services running in each VM, by associating each monitored response time with a timestamp and with the corresponding user transaction. 
This information is used to learn whether/how the response time varies based on the number of user transactions occurring simultaneously, based on the Spearman’s rank correlation coefficient \cite{Sheskin2011_ParametricNonparametricStatisticalProcedures}.
The obtained correlation is then used to check whether a variation in a newly monitored response time corresponds to an expected fluctuation due to an increase/decrease of users’ transactions, or whether it actually denotes a performance anomaly affecting a service.

The above discussed techniques train a baseline model of the application behaviour in an offline learning step, and then employ the trained model to detect anomalies while the application is running. 
This technique works under the assumption that the conditions under which an application is run do not change over time (\eg no new application is deployed on the same VMs) or that what monitored during the training phase is enough to model all possible situations for the application \cite{Xu2017_AdaptiveServiceAPIMonitoring}. 
CloudRanger \cite{29_Wang2018_CloudRanger} tackles the problem from a different perspective, with the ultimate goal of continuously adapting the anomaly detection system to follow the evolution of the application and of the context where it runs. 
CloudRanger \cite{29_Wang2018_CloudRanger} indeed enacts continuous learning on the frontend service of an application, to learn which is its \enquote{expected} response times in the current conditions, \viz based on the monitored response time, load, and throughput.
This is done with the initial, offline training of a baseline model on historically monitored data for the application frontend. 
When the application is running, CloudRanger \cite{29_Wang2018_CloudRanger} applies polynomial regression to monitored KPIs to update the baseline model to reflect the application behaviour in the dynamically changing runtime conditions.
Online anomaly detection is then enacted by checking whether the response time of the application frontend is significantly different from the expected one, \viz if the difference between the expected and monitored response times is higher than a given threshold.

Finally, Hora \cite{27_Pitakrat2018_HoraExtended,28_Pitakrat2016_Hora} tackles the problem of online anomaly detection from yet another perspective, different from all those discussed above.
It indeed combines architectural models with statistical analysis techniques to preemptively determine the occurrence of performance anomalies in a multi-service application. 
Hora exploits SLAstic \cite{vanHoorn2014_SLAstic} to automatically extract a representation of the architectural entities in a multi-service application (\viz application services and nodes used to host them) and the degree to which a component depends on another.
This information is used to create Bayesian networks \cite{Neal1996_BayesianNetwork} modelling the possible propagation of performance anomalies among the services forming an application.
The online anomaly detection is then enacted by associating each application service with a detector, which monitors given KPIs for such service, and which analyses such monitored metrics to detect whether the currently monitored KPIs denote a performance anomaly. 
Since monitored metrics are time series data, detectors exploit autoregressive integrated moving average \cite{Shumway2017_ARIMA} to determine whether performance anomalies affect monitored services. 
Whenever a performance anomaly is detected on a service, Hora \cite{27_Pitakrat2018_HoraExtended,28_Pitakrat2016_Hora} enacts Bayesian inference \cite{Neal1996_BayesianNetwork} to determine whether the anomaly propagates to other services. 
It then returns the set of application services most probably being affected by a performance anomaly.

\subsubsection{Supervised Learning}
ADS \cite{15_Du2018_AnomalyDetectionContainerMicroservices} and PreMiSE \cite{33_Mariani2020_PredictingFailures} also enact online anomaly detection by training a baseline model on monitored KPIs.
They however enact supervised learning (rather than unsupervised learning), by also injecting specific failures in specific services during the training phase.
They indeed label monitored KPIs to denote whether they correspond to normal runs or to runs where specific failures were injected on specific services.
The monitoring is enacted by installing monitoring agents in the VMs used to deploy an application, even if the two techniques consider different types of application deployments and KPIs.
As a result, ADS \cite{15_Du2018_AnomalyDetectionContainerMicroservices} and PreMiSE \cite{33_Mariani2020_PredictingFailures} can model both the normal behaviour of the application services and their anomalous behaviour when failures similar to those injected occur.

ADS \cite{15_Du2018_AnomalyDetectionContainerMicroservices} enables detecting performance anomalies in containerised multi-service applications, given their k8s deployment and modules for injecting failures in their services.
ADS \cite{15_Du2018_AnomalyDetectionContainerMicroservices} considers a predefined set of KPIs for each containerised application service, \viz CPU, memory, and network consumption.
In the training phase, the application is loaded with a workload generator, which is provided by the application operator to simulate end-user requests. 
Multiple runs of the application are simulated, both in failure-free conditions and by exploiting the fault injection modules to inject failures in the application services. 
The monitored KPIs are labelled as pertaining to \enquote{normal} or \enquote{anomalous} runs of the application, in such a way that it can be processed to train a classifier with a supervised machine learning algorithm (\viz nearest neighbour \cite{Altman1992_NearestNeighbor}, random forest \cite{Breiman2001_RandomForests}, na\"ive Bayes \cite{John1995_NaiveBayes}, or support vector machines \cite{Scholkopf1999_SupportVectorLearning}), assuming that no more than one anomaly occurs at the same time.
The trained classifier provides the baseline model for normal/anomalous behaviour of the application, and it is used for classifying application services as experiencing some known performance anomaly based on their newly monitored KPIs.

PreMiSE \cite{33_Mariani2020_PredictingFailures} trains a baseline model for detecting anomalies in multi-service applications, assuming that each service runs in a different VM. 
Monitored KPIs are treated as time series: in an intial, offline learning phase, PreMiSE \cite{33_Mariani2020_PredictingFailures} trains a baseline modelling of the normal, failure-free execution of the application services. 
The baseline model captures temporal relationships in the time series data corresponding to monitored KPIs to model trends and seasonalities, and it applies Granger causality tests \cite{Arnold2007_GrangerMethodsTemporalCausalModelling} to determine whether the correlation among two KPIs is such that one can predict the evolution of the other.
PreMiSE \cite{33_Mariani2020_PredictingFailures} also trains a signature model representing the anomalous behaviour of the application in presence of failures.
This is done by injecting predefined failures (\viz packet loss/corruption, increased network latency, memory leak, and CPU hog) in the VMs running the application services and by monitoring the corresponding changes in KPIs.
The monitored KPIs are then used to train a classifier for detecting the occurrence of the anomalies corresponding to the injected failures, with one out of six supported supervised machine learning algorithms. 
The trained baseline model is then compared against newly monitored KPIs to enact online anomaly detection. 
Once an anomaly is detected, the signature model is used to classify the application service suffering from the detected anomaly and the failure it is suffering from. 

\subsubsection{SLO Check}
\label{sec:detection:monitoring:slo-check}
CauseInfer \cite{40_Chen2020_CauseInfer,48_Chen2014_CauseInfer} and Microscope \cite{06_Lin2018_Microscope,19_Guan2019_MicroscopeDemo} enact online anomaly detection in multi-service applications by monitoring KPIs of the application frontend and comparing such KPIs with the SLOs of the application.
In particular, CauseInfer \cite{40_Chen2020_CauseInfer,48_Chen2014_CauseInfer} exploits a monitoring agent to continuously monitor the response time of the service acting as frontend for an application.
Microscope \cite{06_Lin2018_Microscope,19_Guan2019_MicroscopeDemo} instead relies on the k8s deployment of an application to include monitoring agents (like Prometheus \cite{Prometheus}), throughout which application services expose their response time.
CauseInfer \cite{40_Chen2020_CauseInfer,48_Chen2014_CauseInfer} and Microscope \cite{06_Lin2018_Microscope} and then compare the KPIs monitored on the frontend of an application with that declared in the SLOs of the application. 
Whenever the performance of the application frontend violates the application SLOs, CauseInfer \cite{40_Chen2020_CauseInfer,48_Chen2014_CauseInfer} and Microscope \cite{06_Lin2018_Microscope,19_Guan2019_MicroscopeDemo} consider the whole application as affected by a performance anomaly.


A similar technique is adopted by $\epsilon$-diagnosis \cite{49_Shan2019_epsilonDiagnosis}, which also monitors SLOs on the frontend service of an application to enact online performance anomaly detection.
$\epsilon$-diagnosis \cite{49_Shan2019_epsilonDiagnosis} however focuses on the tail latency of the application frontend, determined in small time windows (\eg one minute or one second), rather than on bigger ones as typically done. 
Given the k8s deployment of the target application, the $\epsilon$-diagnosis system \cite{49_Shan2019_epsilonDiagnosis} enacts online anomaly detection as follows: it partitions the stream of data into small-sized windows, it computes the tail latency for each window, and it then compares the tail latency with the threshold declared in the application SLOs.
If the tail latency for a time window is higher than the given threshold, the whole application is considered to suffer from a performance anomaly in such a time window.

\subsubsection{Heartbeating}
A different, monitoring-based technique to detect anomalous services in multi-service applications is heartbeating.
The latter requires installing a monitorer sending heartbeat messages to the application services, which must reply to such messages within a given timeframe to be considered as fully working.
M-MFSA-HDA \cite{24_Zang2018_FaultDiagnosisMicroservices} adopts this technique, by periodically sending heartbeat messages to the services forming an application.
If any of such services does not reply within a given timeframe, it is considered to suffer from a functional anomaly. 
In addition, to keep a suitable trade-off between the impact of the heartbeat protocol on application performances and the failure detection performances of the heartbeat detection system itself, M-MFSA-HDA \cite{24_Zang2018_FaultDiagnosisMicroservices} self-adaptively changes the heartbeat rate by also monitoring the CPU consumption of the nodes hosting application services, the network load, and the application workload. 

\subsection{Discussion}
\label{sec:detection:discussion}

\begin{table}[tp]
\caption{Classification of anomaly detection techniques, based on their \textit{class} (\viz \textsf{L} for log-based techniques, \textsf{DT} for distributed tracing-based techniques, \textsf{M} for monitoring-based techniques), the applied \textit{method}, the \textit{type} (\viz \textsf{F} for functional anomalies, \textsf{P} for performance anomalies) and \textit{granularity} of detected anomalies (\viz \textsf{A} for application-level anomalies, \textsf{S} service-level anomalies), and the \textit{input} they need to run.}
\label{tab:detection}
\begin{tiny}
\begin{tabular}{%
    >{\centering\arraybackslash}m{.15\textwidth}%
    @{\ }
    >{\centering\arraybackslash}m{.06\textwidth}%
    @{\ }
    >{\centering\arraybackslash}m{.18\textwidth}%
    @{\ }
    >{\centering\arraybackslash}m{.05\textwidth}%
    @{\,}
    >{\centering\arraybackslash}m{.08\textwidth}%
    @{\ }
    >{\centering\arraybackslash}m{.4\textwidth}%
}
    \thickline
        & 
        &
        & \multicolumn{2}{c}{\textbf{Anomaly \ }}
        & 
        \\ 
    \textbf{Reference} 
        & \textbf{Class} 
        & \textbf{Method}
        & \textbf{Type}
        & \textbf{Gran.} 
        & \textbf{Needed Input}
        \\ 
    \thickline
    OASIS \cite{38_Nandi2016_AnomalyDetectionControlFlowGraphMining}
        & \textsf{L} 
        & \cellcolor[HTML]{E7F2F8} unsupervised learning
        & \textsf{F}
        & \textsf{A}
        & previous and runtime logs
        \\
    Jia \etal \cite{31_Jia2017_AnomalyDiagnosisHybridGraphModel}
        & \textsf{L} 
        & \cellcolor[HTML]{E7F2F8} unsupervised learning
        & \textsf{F}, \textsf{P}
        & \textsf{S}
        & previous and runtime logs
        \\
    LogSed \cite{32_Jia2017_LogSed}
        & \textsf{L} 
        & \cellcolor[HTML]{E7F2F8} unsupervised learning
        & \textsf{F}, \textsf{P}
        & \textsf{S}
        & previous and runtime logs
        \\
       \thinline
    TraceAnomaly \cite{16_Liu2020_TraceAnomaly}
        & \textsf{DT} 
        & \cellcolor[HTML]{E7F2F8} unsupervised learning
        & \textsf{F}, \textsf{P}
        & \textsf{A}
        & previous and runtime traces
        \\
    Nedelkoski \etal \cite{35_Nedelkoski2019_AnomalyDetectionTracingData}
        & \textsf{DT}  
        & \cellcolor[HTML]{E7F2F8} unsupervised learning 
        & \textsf{F}, \textsf{P}
        & \textsf{A}
        & previous and runtime traces
        \\
    Jin \etal \cite{22_Jin2020_AnomalyDetectionMicroservicesRPCA}
        & \textsf{DT} 
        & \cellcolor[HTML]{E7F2F8} unsupervised learning 
        & \textsf{P}
        & \textsf{S}
        & previous and runtime traces
        \\
    
    Seer \cite{17_Gan2019_Seer}
        & \textsf{DT} 
        & \cellcolor[HTML]{97CDD8} supervised learning
        & \textsf{P}
        & \textsf{S}
        & previous and runtime traces
        \\
    
    Nedelkoski \etal \cite{34_Nedelkoski2019_AnomalyDetectionDistributedTracing}
        & \textsf{DT}  
        & \cellcolor[HTML]{97CDD8} supervised learning
        & \textsf{P}
        & \textsf{S}
        & previous and runtime traces
        \\
    
    Bogatinovski \etal \cite{36_Bogatinovski2020_SelfSupervisedAnomalyDetection}
        & \textsf{DT} 
        & \cellcolor[HTML]{97CDD8} supervised learning
        & \textsf{P}
        & \textsf{A}
        & previous and runtime traces
        \\
    
    MEPFL \cite{02_Zhou2019_LatentErrorPrediction}
        & \textsf{DT} 
        & \cellcolor[HTML]{97CDD8} supervised learning 
        & \textsf{F}
        & \textsf{S}
        & app deployment, test cases
        \\
    Meng \etal \cite{23_Meng2021_DetectingAnomaliesMicroservices}
        & \textsf{DT}  
        & \cellcolor[HTML]{FFD1C2} trace comparison
        & \textsf{F}, \textsf{P}
        & \textsf{S}
        & previous and runtime traces 
        \\
    Wang \etal \cite{26_Wang2020_WorkflowAwareFaultDiagnosis}
        & \textsf{DT}  
        & \cellcolor[HTML]{FFD1C2} trace comparison
        & \textsf{F}, \textsf{P}
        & \textsf{S}
        & previous and runtime traces
        \\
    Chen \etal \cite{39_Chen2020_MatrixSketchBasedAnomalyDetection}
        & \textsf{DT} 
        & \cellcolor[HTML]{FFD1C2} trace comparison
        & \textsf{P}
        & \textsf{S}
        & previous and runtime traces
        \\
        \thinline
    Gulenko \etal \cite{37_Gulenko2018_DetectingAnomalousBehaviourBlackBoxServices}
        & \textsf{M} 
        & \cellcolor[HTML]{E7F2F8} unsupervised learning
        & \textsf{P}
        & \textsf{S}
        & app deployment, workload generator
        \\
    LOUD \cite{13_Mariani2018_LocalizingFaultsCloudSystems}
        & \textsf{M}  
        & \cellcolor[HTML]{E7F2F8} unsupervised learning
        & \textsf{P}
        & \textsf{S}
        & app deployment, workload generator
        \\
    MicroRCA \cite{04_Wu2020_MicroRCA} 
        & \textsf{M} 
        & \cellcolor[HTML]{E7F2F8} unsupervised learning
        & \textsf{P}
        & \textsf{S}
        & k8s app deployment, workload generator 
        \\ 
    
    Wu \etal \cite{09_Wu2020_MicroserviceDiagnosisDeepLearning}
        & \textsf{M} 
        & \cellcolor[HTML]{E7F2F8} unsupervised learning
        & \textsf{P}
        & \textsf{S}
        & k8s app deployment, workload generator
        \\
    
    DLA \cite{14_Samir2019_DLA}
        & \textsf{M} 
        & \cellcolor[HTML]{E7F2F8} unsupervised learning
        & \textsf{P}
        & \textsf{S}
        & k8s app deployment 
        \\
    
    CloudRanger \cite{29_Wang2018_CloudRanger}
        & \textsf{M} 
        & \cellcolor[HTML]{E7F2F8} unsupervised learning
        & \textsf{P}
        & \textsf{A}
        & app deployment, prev. monitored KPIs
        \\
    
    Hora \cite{27_Pitakrat2018_HoraExtended,28_Pitakrat2016_Hora}
        & \textsf{M} 
        & \cellcolor[HTML]{E7F2F8} unsupervised learning
        & \textsf{P}
        & \textsf{S}
        & app deployment
        \\
    ADS \cite{15_Du2018_AnomalyDetectionContainerMicroservices}
        & \textsf{M} 
        & \cellcolor[HTML]{97CDD8} supervised learning
        & \textsf{P}
        & \textsf{S}
        & k8s app deployment, workload generator, failure injectors
        \\
    PreMiSE \cite{33_Mariani2020_PredictingFailures}
        & \textsf{M} 
        & \cellcolor[HTML]{97CDD8} supervised learning
        & \textsf{F}, \textsf{P}
        & \textsf{S}
        & app deployment, workload generator  
        \\
    
    CauseInfer \cite{40_Chen2020_CauseInfer,48_Chen2014_CauseInfer}
        & \textsf{M} 
        & \cellcolor[HTML]{F3EDCE} SLO check
        & \textsf{P}
        & \textsf{A}
        & app deployment, SLOs
        \\
    
    MicroScope \cite{06_Lin2018_Microscope,19_Guan2019_MicroscopeDemo}
        & \textsf{M} 
        & \cellcolor[HTML]{F3EDCE} SLO check
        & \textsf{P}
        & \textsf{A}
        & k8s app deployment, SLOs
        \\
    
    $\epsilon$-diagnosis \cite{49_Shan2019_epsilonDiagnosis}
        & \textsf{M} 
        & \cellcolor[HTML]{F3EDCE} SLO check
        & \textsf{P}
        & \textsf{A}
        & k8s app deployment, SLOs
        \\
    M-MFSA- HDA \cite{24_Zang2018_FaultDiagnosisMicroservices}
        & \textsf{M} 
        & \cellcolor[HTML]{E6D6C7} heartbeating
        & \textsf{F}
        & \textsf{S}
        & app deployment
        \\
    \thickline
\end{tabular}
\end{tiny}
\end{table}

\Cref{tab:detection} recaps the surveyed anomaly detection techniques by distinguishing their classes, \viz whether they are log-based, distributed tracing-based, or monitoring-based, as well as the method they apply to enact anomaly detection.
The table also classifies the surveyed techniques based on whether they can detect functional or performance anomalies, and on the \enquote{granularity} of detected anomalies, \viz whether they just state that an application is suffering from an anomaly, or whether they distinguish which of its services are suffering from anomalies.
Finally, the table provides insights on the artifacts that must be provided to the surveyed anomaly detection techniques, as input needed to actually enact anomaly detection.

Taking \Cref{tab:detection} as a reference, we hereafter summarise the surveyed anomaly detection techniques (\Cref{sec:detection:discussion:summary}).
We also discuss them under three different perspectives, \viz finding a suitable compromise between type/granularity of detected anomalies and setup costs (\Cref{sec:detection:discussion:type-granularity-costs}), their accuracy, especially in the case of dynamically changing applications (\Cref{sec:detection:discussion:adaptability}), and the need for explainability and countermeasures for detected anomalies (\Cref{sec:detection:discussion:explainability-countermeasures}). 

\subsubsection{Summary}
\label{sec:detection:discussion:summary}
All the surveyed techniques (but for those based on SLO checks \cite{40_Chen2020_CauseInfer,48_Chen2014_CauseInfer,19_Guan2019_MicroscopeDemo,06_Lin2018_Microscope,49_Shan2019_epsilonDiagnosis} or on heartbeating \cite{24_Zang2018_FaultDiagnosisMicroservices}) rely on processing data collected in training runs of the target applications. 
The idea is to use the logs, traces, or KPIs collected in training runs of the application as a baseline against which to compare newly produced logs or traces, or newly monitored KPIs.
Machine learning is the most used approach to train baseline models of the application behaviour: the logs, traces, or KPIs collected during training runs of the application are indeed used to train baseline models with unsupervised learning algorithms
\cite{37_Gulenko2018_DetectingAnomalousBehaviourBlackBoxServices,31_Jia2017_AnomalyDiagnosisHybridGraphModel,32_Jia2017_LogSed,22_Jin2020_AnomalyDetectionMicroservicesRPCA,16_Liu2020_TraceAnomaly,13_Mariani2018_LocalizingFaultsCloudSystems,38_Nandi2016_AnomalyDetectionControlFlowGraphMining,35_Nedelkoski2019_AnomalyDetectionTracingData,28_Pitakrat2016_Hora,27_Pitakrat2018_HoraExtended,14_Samir2019_DLA,29_Wang2018_CloudRanger,04_Wu2020_MicroRCA,09_Wu2020_MicroserviceDiagnosisDeepLearning}, or with supervised learning algorithms if such information is also labelled with the failures that have happened or were injected on services during the training runs \cite{36_Bogatinovski2020_SelfSupervisedAnomalyDetection,15_Du2018_AnomalyDetectionContainerMicroservices,02_Zhou2019_LatentErrorPrediction,34_Nedelkoski2019_AnomalyDetectionDistributedTracing,33_Mariani2020_PredictingFailures,17_Gan2019_Seer}.
The trained baseline models are typically clusters defining the space where newly monitored KPIs should pertain to not be considered anomalous, or classifiers  or deep neural networks to be fed with newly monitored KPIs, logs, or traces to detect whether they denote anomalies.

An alternative approach to machine learning is trace comparison, which is enacted by Chen \etal \cite{39_Chen2020_MatrixSketchBasedAnomalyDetection}, Meng \etal \cite{23_Meng2021_DetectingAnomaliesMicroservices}, and Wang \etal \cite{26_Wang2020_WorkflowAwareFaultDiagnosis}, all based on instrumenting applications to feature distributed tracing.
The idea here is to store the traces that can possibly occur in an application, and then to check whether newly generated traces are similar enough to the stored ones.
This approach however results to be quite time consuming, indeed bringing two out of the three techniques enacting trace comparison to be suited only for offline anomaly detection.
Only Chen \etal \cite{39_Chen2020_MatrixSketchBasedAnomalyDetection} achieves time performances good enough to enact online anomaly detection on large-scale applications, whereas Meng \etal \cite{23_Meng2021_DetectingAnomaliesMicroservices} and Wang \etal \cite{26_Wang2020_WorkflowAwareFaultDiagnosis} explicitly state that their techniques are too time consuming to be used online on medium-/large-scale applications.

\subsubsection{Type and Granularity of Detected Anomalies vs. Setup Costs}
\label{sec:detection:discussion:type-granularity-costs}
The surveyed techniques achieve a different granularity in anomaly detection by applying different methods and requiring different artifacts (\Cref{tab:detection}).
Independently from the required instrumentation and from whether they apply machine learning or trace comparison, the techniques comparing newly collected KPIs, logs, or traces against a baseline turn out to be much finer in the granularity of detected anomalies, if compared with techniques enacting SLO checks.
Indeed, whilst monitoring SLOs on the frontend of an application allows to state that the whole application is suffering from a performance anomaly, the techniques based on machine learning or trace comparison often allow to detect functional or performance anomalies at service-level.
They indeed typically indicate which of its services are suffering from functional or performance anomalies, whilst also providing information on which are the KPIs monitored on a service, the events it logged, or the portion of a trace due to which a service is considered anomalous.
In the case of supervised learning-based techniques \cite{36_Bogatinovski2020_SelfSupervisedAnomalyDetection,15_Du2018_AnomalyDetectionContainerMicroservices,02_Zhou2019_LatentErrorPrediction,34_Nedelkoski2019_AnomalyDetectionDistributedTracing,33_Mariani2020_PredictingFailures,17_Gan2019_Seer}, they also indicate which failures may possibly correspond to the symptoms denoted by the detected anomalies.
All such information is precious for application operators as it eases further investigating on detected anomalies, \eg to check whether they correspond to services truly failing and to determine why they have failed \cite{Brogi2020_FailureCausalities}.

This comes at the price of requiring to execute training runs of multi-service applications to collect logs, traces, or KPIs to train the baseline models against which to compare those produced at runtime. 
In the case of supervised learning-based techniques, it is also required to either provide what needed to automatically inject failures or to label the collected data to distinguish whether it corresponds to normal runs or to runs where specific failures occurred on specific services.
Then, log-based techniques does not require any application instrumentation, as they directly work on the event logs produced by the application services during the training phase and at runtime \cite{38_Nandi2016_AnomalyDetectionControlFlowGraphMining,31_Jia2017_AnomalyDiagnosisHybridGraphModel,32_Jia2017_LogSed}.
Distributed tracing-based techniques instead assume applications to be instrumented to feature distributed tracing \cite{Richardson2018_MicroservicesPatterns}, so as to automatically collect the traces of events produced by their services.
In this case, the traces used to train the baseline models are obtained twofold, \viz either only with failure-free runs of an application  \cite{16_Liu2020_TraceAnomaly,35_Nedelkoski2019_AnomalyDetectionTracingData,22_Jin2020_AnomalyDetectionMicroservicesRPCA,36_Bogatinovski2020_SelfSupervisedAnomalyDetection} or by also considering runs where specific failures occurred on specific services \cite{17_Gan2019_Seer,34_Nedelkoski2019_AnomalyDetectionDistributedTracing,02_Zhou2019_LatentErrorPrediction}.

Monitoring-based techniques work with another different setup: they require deploying monitoring agents together with the application services, and they directly work with the application services themselves (rather than processing their logs or the traces they produce).
Whilst this is the only need for SLO check-based techiniques \cite{40_Chen2020_CauseInfer,48_Chen2014_CauseInfer,49_Shan2019_epsilonDiagnosis} and for DLA \cite{14_Samir2019_DLA}, monitoring-based techniques typically require additional artifacts to generate a baseline modelling of the KPIs monitored on the application.
They indeed either require KPIs already monitored in former production runs of the application \cite{29_Wang2018_CloudRanger}, or workload generators to simulate the traffic that will then be generated by end users \cite{37_Gulenko2018_DetectingAnomalousBehaviourBlackBoxServices,04_Wu2020_MicroRCA,09_Wu2020_MicroserviceDiagnosisDeepLearning,33_Mariani2020_PredictingFailures,15_Du2018_AnomalyDetectionContainerMicroservices}.
ADS \cite{15_Du2018_AnomalyDetectionContainerMicroservices} also requires application operators to provide failure injection modules to inject specific failures in specific services during the training of the baseline model. 
In addition, some of the monitoring-based techniques focus on specific setups for the deployment of an application, \eg requiring their k8s deployment \cite{49_Shan2019_epsilonDiagnosis,04_Wu2020_MicroRCA,09_Wu2020_MicroserviceDiagnosisDeepLearning,14_Samir2019_DLA,06_Lin2018_Microscope,19_Guan2019_MicroscopeDemo} or assuming their VM-based deployment to be such that each service is deployed in a separate VM \cite{33_Mariani2020_PredictingFailures}.


In summary, not all discussed techniques are suitable to be used on existing applications as they are.
They may require to instrument multi-service applications, to adapt their deployment to work with a given technique, or to provide additional artifacts to configure a technique to enact anomaly detection on an application.
Hence, whilst some techniques are known to detect functional/performance anomalies at a finer granularity, this may not be enough to justify the cost for setting them up to work with a given application, \viz to suitably instrument the application or its deployment, or to provide the needed artifacts.
In some other cases, an application may already natively provide what is needed to apply it a given anomaly detection techniques, \eg event logging or distributed tracing.
The choice of which technique to use for enacting anomaly detection on a given application hence consists of finding a suitable trade-off between the desired granularity/type of detected anomalies and the cost for setting up such technique to work with the given application. 
We believe that the classification provided in this paper can provide a first support for application operators needing to take such a choice.

\subsubsection{Accuracy of Anomaly Detection}
\label{sec:detection:discussion:adaptability}
The surveyed techniques (but for those based on SLO checks or heartbeating) share a common assumption, \viz that, at runtime, the application and its services behave similarly to the training runs.
In other words, they assume that applications will feature similar performances, log similar events, or perform similar interactions in a similar order. 
This is however not always the case, since the actual behaviour of multi-service applications also depends on the conditions under which they are executed, which often dynamically changes \cite{Soldani2018_MicroservicesPainsGains}.
For instance, if hosted on the same virtual environments but with different sets of co-hosted applications, multi-service applications may suffer from different contentions on computing resources and hence feature different performances \cite{Medel2018_ResourceManagementPerformanceKubernetes}.
Similar considerations derive from the workload generated by end-users, which often varies with seasonalities \cite{33_Mariani2020_PredictingFailures}.
In addition, multi-service applications can be highly dynamic in their nature, especially if based on microservices: new services can be included to feature new functionalities, or existing services can be replaced or removed as becoming outdated \cite{Soldani2018_MicroservicesPainsGains}.
As a result, all techniques assuming that multi-service applications keep running under the same conditions as in training runs may lower their accuracy if the running conditions of an application change, as they may end up with detecting false positives or with suffering from false negatives.
They may indeed detect functional or performance anomalies in a running application, even if these are not anomalous changes, but rather correspond to the \enquote{new normal behaviour} in the new conditions under which an application runs.
They may instead consider the actual behaviour of an application as normal, even if such behaviour may denote anomalies in the latest running conditions of an application.

The problem of false positives and false negatives is not to be underestimated.
Anomaly detection is indeed enacted as anomalies are possible symptoms of failures.
A detected anomaly on a service hence provides a warning on the fact that such service can have possibly failed, and it may also be used to determine the type of failure affecting such service. 
A false positive, \viz a detected anomaly that is actually not an anomaly, would result the application operator in spending resources and efforts on application services that could have possibly failed, \eg to recover them or to understand why they have failed, even if this was not the case. 
Even worse is the case of false negatives: if some anomalies would not be detected, the symptoms of some failures may not get detected, which in turn means that the application operator would not be warned in case such failures actually occur in a multi-service application.

To mitigate this issue, the baseline used to detect anomalies should be kept up to date by adapting it to the varying conditions under which a multi-service application runs.
A subset of the surveyed techniques explicitly recognise this problem, \eg PreMiSE \cite{33_Mariani2020_PredictingFailures} prescribes to use a huge amount of training data to include as many evolutions of the application as possible, therein including seasonalities of workloads, to mitigate the effects of the \enquote{training problem} as much as possible.
Some of the surveyed techniques also propose solutions to address the training problem by keeping the baseline model up to date.
For instance, DLA \cite{14_Samir2019_DLA}, Seer \cite{17_Gan2019_Seer}, and Wang \etal \cite{26_Wang2020_WorkflowAwareFaultDiagnosis}, among others, propose to periodically re-train the baseline model by considering newly collected data as training data.
They also observe that the re-training period should be tailored as a trade-off between the effectiveness of the trained models and the computational effort needed to re-train the model, as re-training is typically resource and time consuming. 
CloudRanger \cite{29_Wang2018_CloudRanger} instead keeps the baseline continuously updated by enacting continuous learning on the services forming an application.
Among trace comparison-based techniques, Chen \etal \cite{39_Chen2020_MatrixSketchBasedAnomalyDetection} actually try to address the issue with false positives by exploiting matrix sketching to keep track of the entire prior for each service.
They hence compare newly collected information with all what has been observed formerly, both during the offline reference runs and while the application was running in production.

In the case of supervised learning-based techniques \cite{36_Bogatinovski2020_SelfSupervisedAnomalyDetection,15_Du2018_AnomalyDetectionContainerMicroservices,02_Zhou2019_LatentErrorPrediction,34_Nedelkoski2019_AnomalyDetectionDistributedTracing,33_Mariani2020_PredictingFailures,17_Gan2019_Seer}, the false positives and false negatives possibly caused by the training problem impact not only on anomaly detection, but also on the identification of the failure corresponding to a detected anomaly. 
Failure identification is indeed based on the correlation of symptoms, assuming that the same failure happening on the same service would result in a similar set of symptoms. 
Again, the varying conditions under which an application runs, as well as the fact that the service forming an application may dynamically change themselves, may result in the symptoms observed on a service to change over time.
The failures that can possibly affect a service may also change over time, hence requiring to adapt the supervised learning-based techniques not only to the varying conditions under which an application runs, but also to the varying sets of possible failures for a service.
To tackle this issue, Seer \cite{17_Gan2019_Seer} re-trains the baseline modelling of the application by also considering the failures that were observed on the application services while they were running in production.
Seer \cite{17_Gan2019_Seer} is actually the only technique  dealing with the training problem among those enacting supervised learning.
This is mainly because Seer \cite{17_Gan2019_Seer} is designed to be periodically trained on the data monitored while an application is running in production, whereas the other supervised learning-based techniques rely on predefined failures being artificially injected during the training runs.

The above however only constitute mitigation actions, \viz actions allowing to mitigate the false positives/negatives due to the training problem. 
False positives/negatives are indeed a price to pay when detecting anomalies by comparing the runtime behaviour of an application with a baseline modelling of its normal, failure-free behaviour.
A quantitative comparison of the accuracy of the different techniques on given applications in given contexts would further help application operators in this direction, and it will complement the information in this survey in supporting application in choosing the anomaly detection techniques best suited to their needs.
Such a quantitative comparison is however outside of the scope of this survey and a direction for future work.

\subsubsection{Explainability and Countermeasures}
\label{sec:detection:discussion:explainability-countermeasures}
The problem of false positives/negatives is also motivated by a currently open challenge in the field of anomaly detection for multi-service applications, \viz \enquote{explainability}.
This is tightly coupled with the problem of explainable AI \cite{Guidotti2019_Explainability}, given that anomaly detection is mainly enacted with machine learning-based techniques.
If giving explanations for answers given by learned models is generally important, it is even more important in online anomaly detection.
Explanations would indeed allow application operators to directly exclude false positives, or to exploit such explanations, \eg to investigate the root causes for the corresponding failures, as well as to design countermeasures avoiding such failures to happen again \cite{Brogi2020_FailureCausalities}.
Root cause analysis techniques do exist and ---as we show in \Cref{sec:rca}--- they still suffer from the training problem potentially causing false positives or false negatives, as well as from the need for explainability.

Another interesting research direction would be to take advantage of the information processed to detect anomalies to enact countermeasures to isolate (the failures causing) detected anomalies,
\eg by proactively activating circuit breakers/bulkheads \cite{Richardson2018_MicroservicesPatterns} to avoid that a service anomaly propagates to other services.
%
A first attempt in this direction is made 
by Seer \cite{17_Gan2019_Seer}, which notifies the system manager about the anomaly detected on a service and the failure that can have possibly caused it, \eg CPU hog or memory leak. 
With such information, Seer \cite{17_Gan2019_Seer} envisions the system manager to be able to enact policies mitigating the effects of the failure causing a detected anomaly, \eg by providing the node hosting an anomalous service with additional computing resources.

\section{Root Cause Analysis}
\label{sec:rca}
We hereafter survey the existing techniques for determining the possible root causes for anomalies observed in multi-service applications.
Similarly to \Cref{sec:detection}, we first present the class of techniques enacting root cause analysis by directly processing the logs produced by application services (\Cref{sec:rca:service-logs}).
We then present the root cause analysis techniques requiring multi-service applications to feature distributed tracing (\Cref{sec:rca:distributed-tracing}) or to install agents to monitor their services (\Cref{sec:rca:monitoring}).
We finally discuss the surveyed root cause analysis techniques, with the support of a recapping table (\Cref{tab:rca}) and by also highlighting some open challenges in root cause analysis (\Cref{sec:rca:discussion}).

\subsection{Log-based Root Cause Analysis Techniques}
\label{sec:rca:service-logs}
Root cause analysis can be enacted by only considering the logs natively produced by the services forming an application (and without requiring further instrumentation like distributed tracing or monitoring agents).
This is done by processing the application logs to derive a \enquote{causality graph}, whose vertices model application services, and whose directed arcs model that an anomaly in the source service may possibly cause an anomaly in the target service.
The causality graph is then visited to determine a possible root cause for an anomaly observed on a multi-service application.

\subsubsection{{Causality Graph-based Analysis}}
Aggarwal \etal \cite{10_Aggarwal2020_FaultsGoldenSignals} allow to determine the root causes for functional anomalies observed on the frontend of a multi-service application, \viz frontend errors that users face when the application fails.
Aggarwal \etal \cite{10_Aggarwal2020_FaultsGoldenSignals} consider the subset of application services including the frontend and the services that logged error events.
They model the logs of considered services as multivariate time series, and they compute Granger causality tests \cite{Arnold2007_GrangerMethodsTemporalCausalModelling} on such time series to determine 
causal dependencies among the errors logged by the corresponding services.
The obtained cause-effect relations are used to refine the topology specified in an input application specification, so as to derive a causality graph, \viz a graph whose nodes model application services, and where an arc from a service to another that the errors logged by the source service can have caused those logged by the target service.
The services that, according to the causality graph, can have caused directly or indirectly the error observed on the application frontend are considered as the possible root causes for the corresponding faults.
The candidate services are scored by visiting the causality graph with the random walk-based algorithm proposed by MonitorRank \cite{21_Kim2013_RootCauseDetectionSOA} (which we present in \Cref{sec:rca:distributed-tracing:topology}).
Aggarwal \etal \cite{10_Aggarwal2020_FaultsGoldenSignals} then return the highest scored candidate, which is considered the most probable root cause for the observed functional anomaly.

\subsection{Distributed Tracing-based Root Cause Analysis Techniques}
\label{sec:rca:distributed-tracing}
The information available in distributed traces can be exploited to support application operators in determining the possible root causes for anomalies observed in their applications.
In this perspective, the most basic technique consists of providing a systematic methodology to visually compare traces, so that application operators can manually determine what went wrong in the \enquote{anomalous traces}, \viz the traces corresponding to observed anomalies.

To automate the root cause analysis, the alternatives are twofold.
An alternative is the directly analysing traces to detect which services experienced anomalies, which may have possibly caused the observed anomalies.
The other possibility is to automatically determine the topology of an application (\viz a directed graph whose vertices model the application services and whose directed arcs model the service interactions) from the available traces, and to use such topology to drive the analysis for determining the possible root causes for anomalies observed on an application.

\subsubsection{{Visualization-based Analysis}}
Zhou \etal \cite{01_Zhou2021_FaultAnalysisDebuggingMicroservices} and GMTA \cite{45_Guo2020_GMTA} allow application operators to manually determine the possible root causes for application-level anomalies, based on a trace comparison methodology to be enacted with the support of a trace visualisation tool.
Zhou \etal \cite{01_Zhou2021_FaultAnalysisDebuggingMicroservices} propose to use ShiViz \cite{Beschastnikh2016_ShiViz} to visualise the traces produced by a multi-service application as interactive time-space diagrams.
This allows to determine the root causes for the traces corresponding to some anomaly observed in the application frontend by means of pairwise comparison of traces.
The idea is that the root causes for an anomalous trace are contained in a portion of such trace that is different from what contained in a successful trace for the same task, whilst shared with other anomalous traces. 
Zhou \etal \cite{01_Zhou2021_FaultAnalysisDebuggingMicroservices} hence propose to consider the traces corresponding to the observed anomaly and a set of reference, successful traces for the same set of tasks. 
They then propose to compare a successful trace and an anomalous trace corresponding to the same task. 
This allows obtaining a set of so-called \enquote{diff ranges}, each of which corresponds to multiple consecutive events that are different between the two traces. 
Zhou \etal \cite{01_Zhou2021_FaultAnalysisDebuggingMicroservices} propose to repeat this process for any pair of successful and anomalous traces for the same task, by always restricting the set of \enquote{diff ranges} to those that are shared among all analysed traces. 
The \enquote{diff ranges} obtained at the end of the process are identified as the possible root causes for the observed anomaly. 

GMTA \cite{45_Guo2020_GMTA} provides a framework to collect, process, and visualise traces, which can be used to determine the root causes for functional anomalies observed on a multi-service application.
GMTA \cite{45_Guo2020_GMTA} collects the traces produced by the application services, and it automatically process them to determine interaction \enquote{paths} among services.
A path is an abstraction for traces, which only models the service interactions in a trace and their order with a tree-like structure.
GMTA \cite{45_Guo2020_GMTA} also groups paths based on their corresponding \enquote{business flow}, defined by application operators to indicate which operations could be invoked to perform a task and in which order.
GMTA \cite{45_Guo2020_GMTA} then allows to manually enact root cause analysis by visually comparing the paths corresponding to traces of successful executions of a given business flow with that corresponding to the anomalous trace where the functional anomaly was observed. 
This allows to determine whether the paths diverge and, if this is the case, which events possibly caused the paths to diverge.
In addition, if the anomalous traces include error events, GMTA \cite{45_Guo2020_GMTA} visually displays such errors together with the corresponding error propagation chains. 
This further allows to determine the services that first logged error events in the anomalous trace, hence possibly causing the observed anomaly. 

Both Zhou \etal \cite{01_Zhou2021_FaultAnalysisDebuggingMicroservices} and GMTA \cite{45_Guo2020_GMTA} allow to determine the root causes for anomalies observed on the frontend of an application, \viz the events that caused the trace to behave anomalously, and their corresponding services. 
They also highlight that by looking at the root cause events, or by inspecting the events logged by the corresponding services close in time to such events, application operators can determine the reasons for such services to cause the observed anomaly, \viz whether they failed internally, because of some unexpected answer from the services they invoked, or because of environmental reasons (\eg lack of available computing resources).

\subsubsection{{Direct Analysis}}
CloudDiag \cite{41_Mi2013_PerformanceDiagnosisCloud} and TraceAnomaly \cite{16_Liu2020_TraceAnomaly} determine the root causes of anomalies observed on the frontend of an application
{%
by directly analysing the response times in the service interactions involved in collected traces.
Their aim is determining other anomalies that may have possibly caused the observed one. 
CloudDiag \cite{41_Mi2013_PerformanceDiagnosisCloud} and TraceAnomaly \cite{16_Liu2020_TraceAnomaly}  however apply two different methods for analysing such response times.
}

CloudDiag \cite{41_Mi2013_PerformanceDiagnosisCloud} allows application operators to manually trigger the root cause analysis when a performance anomaly is observed on the application frontend.
It then processes collected traces to determine \enquote{anomalous groups}, \viz groups of traces sharing the same service call tree and where the coefficient of variation \cite{Neil2010_CoefficientVariation} of response times in service interactions is above a given threshold. 
Anomalous groups are processed to identify the services that most contributed to the anomaly, based on Robust Principal Component Analysis (RPCA). 
When a matrix is corrupted by gross sparse errors, RPCA can extract the columns where the errors appear \cite{Candes2009_RobustPrincipalComponentAnalysis}.
CloudDiag \cite{41_Mi2013_PerformanceDiagnosisCloud} represents the traces in a category as a matrix, and it exploits RPCA to determine the error columns corresponding to services with anomalous response times.
Such services constitute the possible root causes for the observed anomaly, which are returned by CloudDiag \cite{41_Mi2013_PerformanceDiagnosisCloud}.
Root causing services are ranked based on the number of times they appeared as anomalous in anomalous categories: the larger the number of times is, the more possible is that a service caused the observed anomaly.

TraceAnomaly \cite{16_Liu2020_TraceAnomaly} instead determines the possible root causes for the functional and performance anomalies it detects while a multi-service application is running.
As discussed in \Cref{sec:detection:distributed-tracing:unsupervised-learning}, functional anomalies are detected when previously unseen call paths occur in a trace.
The root causes for functional anomalies hence obviously coincides with the services starting the first interaction of the previously unseen call paths.
Performance anomalies are instead detected by means of a deep Bayesian neural network, which classifies newly produced traces as normal/anomalous, without providing insights on the services which caused a trace to get classified as anomalous.
TraceAnomaly \cite{16_Liu2020_TraceAnomaly} hence enacts a root cause analysis by correlating the response times of service interactions in the anomalous trace with the average response times in normal conditions: each service interaction whose response time significantly deviates from the average response time is considered as anomalous.
This results in one or more sequences of anomalous service interactions.
The root causes for the detected performance anomaly are extracted from such anomalous service interactions sequences, by picking the last invoked services as possible root causes.

\subsubsection{{Topology Graph-based Analysis}}
\label{sec:rca:distributed-tracing:topology}
The approach of building and processing topology graphs for determining the possible root causes of performance anomalies is adopted by MonitorRank \cite{21_Kim2013_RootCauseDetectionSOA} and MicroHECL \cite{47_Liu2021_MicroHECL}.
They indeed both rely on services to produce interaction traces, including the start and end times of service interactions, their source and target services, performance metrics (\viz latency, error count, and throughput), and the unique identifier of the corresponding user request. 
This information is then processed to reconstruct service invocation chains for the same user request, which are then combined to build the application topology.
They however differ because they consider different time slices for building topologies, and since MonitorRank \cite{21_Kim2013_RootCauseDetectionSOA} and MicroHECL \cite{47_Liu2021_MicroHECL} exploit topologies to determine the root causes of application-level and service-level anomalies, respectively.
{MonitorRank \cite{21_Kim2013_RootCauseDetectionSOA} and MicroHECL \cite{47_Liu2021_MicroHECL} also differ in the method applied to enact root cause analysis: MonitorRank \cite{21_Kim2013_RootCauseDetectionSOA} performs a random walk on the topology graph, whereas MicroHECL explores it through a breadth-first search (BFS).}

MonitorRank \cite{21_Kim2013_RootCauseDetectionSOA} enacts batch processing to periodically process the traces produced by application services to generate a topology graph for each time period. 
Based on such graphs, MonitorRank \cite{21_Kim2013_RootCauseDetectionSOA} allows to determine the possible root causes of a performance anomaly observed on the application frontend in a given time period.
This is done by running the personalised PageRank algorithm \cite{Glen2003_PersonalizedPageRank} on the topology graph corresponding to the time period of the observed anomaly to determine its root causes.
MonitorRank \cite{21_Kim2013_RootCauseDetectionSOA} starts from the frontend service, and it visits the graph by performing a random walk with a fixed number of random steps.
At each step, the next service to visit is randomly picked in the neighbourhood of the service under visit, \viz among the service under visit, those it calls, and those calling it. 
The pickup probability of each neighbour is proportional to its relevance to the anomaly to be explained, computed based on the correlation between its performance metrics and those of the frontend service (with such metrics available in the traces produced in the considered time period). 
MonitorRank \cite{21_Kim2013_RootCauseDetectionSOA} then returns the list of visited services, ranked by the number of times they have been visited.
The idea is that, given that services are visited based on the correlation of their performances with those of the application frontend, the more are the visits to a service, the more such service can explain the performance anomaly observed on the application frontend. 

MicroHECL \cite{47_Liu2021_MicroHECL} instead builds the topology based on the traces produced in a time window of a given size, which ends at the time when a service experienced the performance anomaly whose root causes are to be determined.
Based on such topology, MicroHECL \cite{47_Liu2021_MicroHECL} determines the anomaly propagation chains that could have led to the observed anomaly.
This is done by starting from the service where the anomaly was observed and by iteratively extending the anomaly propagation chains along the possible anomaly propagation directions, \viz from callee to caller for latency/error count anomalies, and from caller to callee for throughput anomalies. 
At each iteration, the services that can be reached through anomaly propagation directions are included in the anomaly propagation chains if they experienced corresponding anomalies. 
This is checked by applying formerly trained machine learning models for offline detection of anomalies on each service, \viz support vector machines \cite{Scholkopf1999_SupportVectorLearning} for latency anomalies, random forests \cite{Breiman2001_RandomForests} for error count anomalies, and 3-sigma rule  \cite{Chandola2009_SurveyAnomalyDetection} for throughput anomalies. 
When anomaly propagation chains cannot be further extended, the services at the end of each determined propagation chain are considered as possible root causes for the initial anomalous service.
MicroHECL \cite{47_Liu2021_MicroHECL} finally ranks the possible root causes, based on the Pearson correlation between their performance metrics and those of the service where the performance anomaly was initially observed.

\subsection{Monitoring-based Root Cause Analysis Techniques}
\label{sec:rca:monitoring}
Root cause analysis is also enacted by processing the KPIs monitored on the services forming an application, throughout monitoring agents installed alongside the application services.
The techniques enacting such a kind of analysis can be further partitioned in three sub-classes, based on the method they apply to determine the possible root causes for an anomaly observed on a service.
A possibility is to 
{%
directly process monitored KPIs to determine other services that experienced anomalies, which may have possibly caused that actually observed.
}
Other alternatives rely on topology/causality graphs modelling the dependencies among the services in an application, as they visit such graphs to determine which services may have possibly caused the observed anomaly.

\subsubsection{{Direct Analysis}}
When an anomaly is detected on the frontend of a multi-service application, its possible root causes can be determined by identifying the KPIs monitored on the application services that were anomalous in parallel with the frontend anomaly.
The idea is indeed that the anomalous KPIs ---and their corresponding services--- can have possibly caused the detected frontend anomaly.
This technique is actually enacted by $\epsilon$-diagnosis \cite{49_Shan2019_epsilonDiagnosis} to automatically determine the possible root causes of the performance anomalies it detects while the application is running (\Cref{sec:detection:monitoring:slo-check}).
$\epsilon$-diagnosis \cite{49_Shan2019_epsilonDiagnosis} considers the time series of KPIs monitored on each container running the application services, from which it extracts two same-sized samples. 
A sample corresponds to the time period where the frontend anomaly was detected, while the other sample corresponding to a time period where no anomaly was detected.
To determine whether a KPI can have determined the detected anomaly, the similarity between the two samples is computed: if the similarity between the two samples is beyond a given threshold, this means that no change on the KPI occurred while the frontend anomaly was monitored, hence meaning that such KPI cannot have caused such anomaly. 
Instead, if the similarity is below the given threshold, the two samples are diverse enough to consider the change in such KPI as anomalous and possibly causing the frontend anomaly.
The list of anomalous KPIs, together with the corresponding services, is returned by $\epsilon$-diagnosis \cite{49_Shan2019_epsilonDiagnosis} as the set of possible root causes.

Wang \etal \cite{20_Wang2020_RootCauseMetricLocation}, PAL \cite{42_Nguyen2011_PAL}, and FChain \cite{43_Nguyen2013_FChain} also exploit offline detection of anomalous KPIs to determine the possible root causes for performance anomalies detected on the frontend of a multi-service application, by however relying on external monitoring tools to detect such anomalies.
The root cause analysis proposed by Wang \etal \cite{20_Wang2020_RootCauseMetricLocation} processes both the logs natively produced by the services forming an application and the time series of KPIs monitored on such services by devoted monitoring agents.
They first train a long-short term memory neural network for each service, which models its baseline behaviour in failure-free runs.
When an anomaly is observed on the application, each trained neural network processes the logs produced by the corresponding service in the time interval when the anomaly was observed.
The outputs of all trained neural networks are combined to obtain a time series of anomaly degrees for the application.
Wang \etal \cite{20_Wang2020_RootCauseMetricLocation} then apply mutual information \cite{Peng2005_MutualInformation} to determine the dependency of the anomaly degree on the KPIs monitored on application services in the same interval. 
The obtained correlation values are used to provide the list of KPIs that may have possibly caused the observed application anomaly, together with the services on which they have been monitored. 
The list of root causes is sorted by decreasing correlation value, with the goal of first showing the services' KPIs having higher probability to have caused the observed anomaly.

PAL \cite{42_Nguyen2011_PAL} and FChain \cite{43_Nguyen2013_FChain} propose two similar techniques to determine the possible root causes for an application-level performance anomaly, both based on the analysis of system-level KPIs monitored on application services (\viz CPU usage, free memory, and network traffic).
Their assumption is indeed that performance anomalies are manifested also as observable changes in one or multiple system-level KPIs.
When a performance anomaly is detected on the application frontend, PAL \cite{42_Nguyen2011_PAL} and FChain \cite{43_Nguyen2013_FChain} examine each system-level KPI monitored on each application service in a look-back window of a given size. 
Each KPI is analysed with a combination of CUSUM (Cumulative Sum) charts and bootstrapping \cite{Basseville1993_DetectionAbruptChanges}.
CUSUM charts are used to measure the magnitude of changes for the monitored KPI values, both in the original sequence of monitored values and in \enquote{bootstraps}, \viz permutations of the monitored KPI values obtained by randomly reordering them. 
If the original sequence of monitored KPI values has a magnitude of change higher than most bootstraps, it is considered anomalous and the time when the anomaly started is identified based on the CUSUM charts of the original sequence.
PAL \cite{42_Nguyen2011_PAL} and FChain \cite{43_Nguyen2013_FChain} then check whether anomalies affected all application services. 
If this is the case, the anomaly is classified as due to external reasons (\eg workload spikes). 
Instead, if only a subset of application services is affected by anomalies, such services are considered as the possible root causes for the frontend anomaly.
PAL \cite{42_Nguyen2011_PAL} and FChain \cite{43_Nguyen2013_FChain} return the possible root causing services sorted by chronologically ordering the start times of their anomalies.
The idea is indeed that the earliest anomalies may have propagated from their corresponding services to other services up to the application frontend, hence most probably being the root causes for the observed frontend anomaly.

\subsubsection{{Topology Graph-based Analysis}}
MicroRCA \cite{04_Wu2020_MicroRCA}, Wu \etal \cite{09_Wu2020_MicroserviceDiagnosisDeepLearning}, Sieve \cite{30_Thalheim2017_Sieve}, and Brand\'on \etal \cite{11_Brandon2020_GraphBasedRootCauseAnalysis} automatically reconstruct the graph modelling the topology of a running application, which they then use to drive the analysis of the possible root causes for observed anomalies.
{%
DLA \cite{14_Samir2019_DLA} also exploits a topology graph to identify the possible root causes for observed anomalies, but rather relying on the application operator to provide such graph.
The above techniques however differ in the methods applied over topology graphs to determine the possible root causes of observed anomalies, either enacting a random walk over topology graph or applying other analysis methods. 
}

\paragraph{Random Walk}
MicroRCA \cite{04_Wu2020_MicroRCA} considers the status of the Kubernetes deployment of an application when an anomaly was detected, \viz which containerised service were running, on which nodes they were hosted, and the interactions that occurred among services. 
MicroRCA \cite{04_Wu2020_MicroRCA} exploits this information to generate a topology graph, whose vertices model the running application services and the node used to host them, and whose oriented arcs model service interactions or service hosting. 
The topology graph is enriched by also associating each vertex with the time series of KPIs monitored on the corresponding service or node.
MicroRCA \cite{04_Wu2020_MicroRCA} then extracts an \enquote{anomalous subgraph} from the topology graph.
The anomalous subgraph is obtained by extracting the vertices corresponding to the services where anomalies were detected, by including the vertices and arcs corresponding the interactions to/from the anomalous services, and by including other vertices and arcs from the topology graph so as to obtain a connected subgraph.
The anomalous subgraph constitutes the search space where MicroRCA \cite{04_Wu2020_MicroRCA} looks for the services that may have possibly caused the detected anomalies.
The actual search for the root causing services is enacted by adapting the random walk-based search proposed in MonitorRank \cite{21_Kim2013_RootCauseDetectionSOA} (\Cref{sec:rca:distributed-tracing:topology}) to work with the information modelled by the anomalous subgraph.

Wu \etal \cite{09_Wu2020_MicroserviceDiagnosisDeepLearning} refine MicroRCA \cite{09_Wu2020_MicroserviceDiagnosisDeepLearning} to determine the services' KPIs that may have possibly caused some anomalies to be detected on some services.
They reuse the topology-driven search proposed by MicroRCA \cite{04_Wu2020_MicroRCA} to determine the services that can have possibly caused the detected anomalies.
In addition, Wu \etal \cite{09_Wu2020_MicroserviceDiagnosisDeepLearning} process the KPIs monitored on root causing services with the autoencoder \cite{Goodfellow2016_DeepLearningBook}, \viz a neural network that learns to encode input values and to suitably reconstruct them from their encoding.
Wu \etal \cite{09_Wu2020_MicroserviceDiagnosisDeepLearning} first train an autoencoder with values for such KPIs monitored while no anomaly was affecting the application.
They then feed the trained autoencoder with the KPI values monitored on the application services when the considered anomalies were detected.
The KPIs whose values are not well-reconstructed, together with their corresponding services, are considered as the possible root causes for the detected anomalies.

\paragraph{Other Methods}
Sieve \cite{30_Thalheim2017_Sieve} and Brand\'on \etal \cite{11_Brandon2020_GraphBasedRootCauseAnalysis} monitor network system calls to collect information on service interactions, which they then exploit to automatically reconstruct the topology of a running application.
When a performance anomaly is observed on the application frontend, Sieve \cite{30_Thalheim2017_Sieve} exploits such topology to drive the analysis of the possible root causes for the observed anomaly.
It first reduces the dimensionality of to-be-considered KPIs, by removing those whose variance is too small to provide some statistical value, and by clustering the KPIs monitored on each service so as to consider only one representative KPI for each cluster, \viz that closest to the centroid of the cluster.
Sieve \cite{30_Thalheim2017_Sieve} then explores the possibilities of a service’s representative KPIs to influence other services’ KPIs using a pairwise comparison method: each representative KPI of each service is compared with each representative KPIs of another service. 
The choice of which services to compare is driven by the application topology: 
Sieve \cite{30_Thalheim2017_Sieve} indeed enacts a pairwise comparison of KPIs monitored only on interacting services, by exploiting Granger causality tests \cite{Arnold2007_GrangerMethodsTemporalCausalModelling} to determine whether changes in the values monitored for a service's KPI were caused by changes in those monitored for another service's KPI.
The obtained cause-effect relations are used to only keep the KPIs whose changes were not caused by other services.
Such KPIs, together with their corresponding services, are considered as the possible root causes for the observed anomaly.

Brand\'on \etal \cite{11_Brandon2020_GraphBasedRootCauseAnalysis} enrich the derived topology by associating each service with its monitored KPIs and with the possibility of marking services as anomalous, by however relying on existing anomaly detection techniques to actually detect anomalous services.
Whenever the root cause for an anomaly in a service is looked for, the proposed system extracts subgraph modelling the neighbourhood for an anomalous service, \viz it includes the vertices close to that modelling the anomalous service up to a given distance.
The anomalous subgraph is compared with \enquote{anomaly graph patterns} provided by the application operator, \viz with graphs modelling already troubleshooted anomalies and whose root causing service is known.
If the similarity between an anomalous subgraph and an anomaly graph pattern is above a given threshold, the corresponding root causing service is included among the possible root causes for the considered anomaly.
If multiple root causing services are detected, they are ranked by similarity between the corresponding anomaly graph patterns and the anomalous subgraph.

DLA \cite{14_Samir2019_DLA} instead starts from the application topology provided by the application operator, which represents the services forming an application, the containers used to run them, and the VMs where the containers are hosted, together with the communication/hosting relationships between them.
DLA \cite{14_Samir2019_DLA} transforms the input topology into a Hiearchical Hidden Markov Model (HHMM) \cite{Capp2010_HiddenMarkovModels} by automatically determining the probabilities of an anomaly affecting a service, container, or VM to be caused by the components it relates to, \viz by the components it interacts with, or by those used to host it.
Such probabilities are automatically determined when an anomaly is detected, by processing the latest monitored KPIs (\viz response time and CPU, memory, and network consumption) with a combination of the Baum-Welch and Viterbi algorithms \cite{Capp2010_HiddenMarkovModels}.
The possible root causes for the detected anomaly are then determined by exploiting the obtained HHMM to compute the likelihood of the anomaly to be generated by KPI anomalies monitored at container- or VM-level. More precisely, DLA computes the path in the HHMM that has the highest probability to have caused the anomaly observed on a service. 
The obtained path is then used to elicit the KPIs (and corresponding components) that most probably caused the observed anomaly.

\subsubsection{{Causality Graph-based Analysis}}
Various existing techniques determine the possible root causes for an anomaly observed on some application service by suitably visiting an automatically derived causality graph \cite{40_Chen2020_CauseInfer,48_Chen2014_CauseInfer,06_Lin2018_Microscope,19_Guan2019_MicroscopeDemo,44_Lin2018_FacGraph,07_Ma2019_MSRank,25_Ma2020_MSRankExtended,08_Ma2020_AutoMAP,29_Wang2018_CloudRanger}.
The vertices in the causality graph typically model application services, with each oriented arc indicating that the performance of the source service depend on that of the target service.
Causality graph are typically built by exploiting the well-known PC-algorithm \cite{Spirtes2000_CausationPredictionSearchPCAlgoritm}.
More precisely, the causality graph is built by starting from a complete and undirected graph, which is refined by dropping and orienting arcs to effectively model causal dependencies among the KPIs monitored on services. 
The refinement is enacted by pairwise testing conditional independence between the KPIs monitored on different services: if the KPIs monitored on two services result to be conditionally independent, the arc between the services is removed. 
The remaining arcs are then oriented based on the structure of the graph.

{The existing causality graph-based techniques however differ in the methods applied to process causality graphs to determine the possible root causes for observed anomalies.
The most common methods are visiting the causality graph through a BFS or a random walk, but there exist also techniques enacting other causality graph-based analysis methods.} 

\paragraph{{BFS}}
CauseInfer \cite{40_Chen2020_CauseInfer,48_Chen2014_CauseInfer} and Microscope \cite{06_Lin2018_Microscope,19_Guan2019_MicroscopeDemo} determine the possible root causes for the response time anomalies they detect on the frontend of a multi-service application.
They rely on monitoring agents to collect information on service interactions and their response time, with CauseInfer \cite{40_Chen2020_CauseInfer,48_Chen2014_CauseInfer} installing them in the nodes where the application services run, whilst Microscope \cite{06_Lin2018_Microscope,19_Guan2019_MicroscopeDemo} assuming monitoring agents to be included in the k8s deployment of an application.
The monitored information is processed with the PC-algorithm to build a causality graph, by exploiting $G^2$ cross-entropy \cite{Spirtes2000_CausationPredictionSearchPCAlgoritm} to test conditional independence between the response times monitored on different services.
The causality graph is enriched by also including arcs modelling the dependency of each service with the services it invokes.
On this basis, CauseInfer \cite{40_Chen2020_CauseInfer} and Microscope \cite{06_Lin2018_Microscope,19_Guan2019_MicroscopeDemo} enact root cause inference by recursively visiting the obtained causality graph, starting from the application frontend.
For each visited service, they consider the services on which it depends: if all such services were not affected by response time anomalies, the currently visited service is considered a possible root cause.
Otherwise, CauseInfer \cite{40_Chen2020_CauseInfer,48_Chen2014_CauseInfer} and Microscope \cite{06_Lin2018_Microscope,19_Guan2019_MicroscopeDemo} recursively visit all services that were affected by response time anomalies to find the possible root causes for the anomaly observed on the currently visited service.
To determine anomalies in the response times of visited services, CauseInfer \cite{40_Chen2020_CauseInfer,48_Chen2014_CauseInfer} computes an anomaly score based on the cumulative sum statistics on the history of monitored response times.
Microscope \cite{06_Lin2018_Microscope,19_Guan2019_MicroscopeDemo} instead exploits standard deviation-based heuristics to determine whether the latest monitored response times were anomalous.
When all possible root causes have been identified, CauseInfer \cite{40_Chen2020_CauseInfer,48_Chen2014_CauseInfer} ranks them based on their anomaly score, whilst Microscope \cite{06_Lin2018_Microscope,19_Guan2019_MicroscopeDemo} ranks them based on the Pearson correlation between their response times and that of the application frontend.

{%
Qiu \etal \cite{46_Qiu2020_CausalityMiningKnowledgeGraph} instead consider service-level performance anomalies, \viz they aim at determining the possible root causes for an anomaly observed in the KPIs monitored on any service in an application (rather than on the application frontend).
To determine such root causes, they exploit the PC-algorithm \cite{Spirtes2000_CausationPredictionSearchPCAlgoritm} to build causality graphs whose vertices correspond to KPIs monitored on application services (rather than to the services themselves), \viz their response times, error count, queries per second, and resource consumption.
Qiu \etal \cite{46_Qiu2020_CausalityMiningKnowledgeGraph} also enrich the causality graph by including arcs among KPIs also when a dependency between the corresponding services is indicated in the application specification provided as input by the application operator.
The arcs in the causality graph are also weighted, with each arc's weight indicating how much the source KPI influences the target KPI. 
The weight of each arc is computed by determining the Pearson correlation between the sequence of significant changes in the source and target KPIs, with significant changes being determined by processing the corresponding time series of monitored KPI values with the solution proposed by Luo \etal \cite{Luo2014_CorrelatingEventsIncidentDiagnosis}.
The root cause analysis is then enacted with a BFS in the causality graph, starting from the KPI whose anomaly was observed.
This allows to determine all possible paths outgoing from the anomalous KPI in the causality graph, which all correspond to possible causes for the anomaly observed on such KPI. 
The paths are then sorted based on the sum of the weights associated to the edges in the path by prioritizing shorter paths in the case of paths whose sum is equivalent. 
In this way, Qiu \etal \cite{46_Qiu2020_CausalityMiningKnowledgeGraph} prioritize the paths including the KPIs whose changes most probably caused the observed anomaly.
}

\paragraph{{Random Walk}}
CloudRanger \cite{29_Wang2018_CloudRanger}, MS-Rank \cite{07_Ma2019_MSRank,25_Ma2020_MSRankExtended}, and AutoMAP \cite{08_Ma2020_AutoMAP} exploit the PC-algorithm to build a causality graph by processing the KPIs monitored on application services.
As for KPIs, CloudRanger \cite{29_Wang2018_CloudRanger} considers response time, throughput, and power.
Such KPIs are also considered by MS-Rank \cite{07_Ma2019_MSRank,25_Ma2020_MSRankExtended} and AutoMAP \cite{08_Ma2020_AutoMAP}, in addition to the services' availability and resource consumption.
CloudRanger \cite{29_Wang2018_CloudRanger}, MS-Rank \cite{07_Ma2019_MSRank,25_Ma2020_MSRankExtended}, and AutoMAP \cite{08_Ma2020_AutoMAP} however differ from CauseInfer \cite{40_Chen2020_CauseInfer} and Microscope \cite{06_Lin2018_Microscope,19_Guan2019_MicroscopeDemo} because they exploit $d$-separation \cite{Cohen2005_DSeparation} (rather than $G^2$ cross entropy) to test conditional independence while building the causality graph with the PC-algorithm \cite{Spirtes2000_CausationPredictionSearchPCAlgoritm}.
They also differ in the way they enact root cause analysis on obtained causality graphs:
a causality graph indeed provides the search space where CloudRanger \cite{29_Wang2018_CloudRanger}, MS-Rank \cite{07_Ma2019_MSRank,25_Ma2020_MSRankExtended}, and AutoMAP \cite{08_Ma2020_AutoMAP} perform a random walk to determine the possible root causes for a performance anomaly observed on the application frontend. 
The random walk starts from the application frontend and it consists of repeating $n$ times (with $n$ equal to the number of services in an application) a random step to visit one of the services in the neighbourhood of the currently visited service, \viz the set including the currently visited service, together with the services that can be reached by traversing forward/backward arcs connected to the currently visited service.
At each iteration, the probability of visiting a service in the neighbourhood is proportional to the correlation of its KPIs with those of the application frontend. 
CloudRanger \cite{29_Wang2018_CloudRanger}, MS-Rank \cite{07_Ma2019_MSRank,25_Ma2020_MSRankExtended}, and AutoMAP \cite{08_Ma2020_AutoMAP} rank the application services based on how many times they have been visited, considering that most visited services constitute the most probable root causes for the anomaly observed on the frontend.

{%
Differently from the above discussed techniques \cite{29_Wang2018_CloudRanger,07_Ma2019_MSRank,08_Ma2020_AutoMAP,25_Ma2020_MSRankExtended}, MicroCause \cite{05_Meng2020_RootCausesMicroservices} determines the root causes of service-level performance anomalies by exploiting the PC-algorithm \cite{Spirtes2000_CausationPredictionSearchPCAlgoritm} to build a causality graph whose vertices correspond to KPIs monitored on application services (rather than to the services themselves).
In particular, MicroCause \cite{05_Meng2020_RootCausesMicroservices} considers the response time monitored on application services, their error count, queries per second, and resource consumption.
MicroCause \cite{05_Meng2020_RootCausesMicroservices} then determines the KPIs that experienced anomalies and the time when their anomaly started by exploiting the SPOT algorithm \cite{Siffer2017_SPOTAlgorithm}.
MicroCause \cite{Siffer2017_SPOTAlgorithm} also computes an anomaly score for each KPI, essentially by measuring the magnitude of its anomalous changes.
This information is then used to drive a random walk over the causality graph to determine the possible root causes for an anomalous KPI being observed on a service.
MicroCause \cite{05_Meng2020_RootCausesMicroservices} starts from the anomalous KPI and repeats a given number of random steps, each consisting of staying in the currently visited KPI or randomly visiting a KPI that can be reached by traversing forward/backward arcs connected to the currently visited KPI.
At each step, the probability of visiting a KPI is based on the time and score of its anomaly, if any, and on the correlation of its values with those of the anomalous KPI whose root causes are being searched.
The KPIs visited during the random walk constitute the set of possible root causes returned by MicroCause \cite{05_Meng2020_RootCausesMicroservices}, ranked based on their anomaly time and score.
}

\paragraph{{Other Methods}}
FacGraph \cite{44_Lin2018_FacGraph} exploits the PC-algorithm \cite{Spirtes2000_CausationPredictionSearchPCAlgoritm} with $d$-separation \cite{Cohen2005_DSeparation} to build causality graphs, whilst also assuming the same frontend anomaly to be observed in multiple time intervals.
It indeed exploits the PC-algorithm to build multiple causality graphs, each built on the latency and throughput monitored on application services during the different time intervals when the frontend anomaly was observed. 
FacGraph \cite{44_Lin2018_FacGraph} then searches for anomaly propagation subgraphs in the obtained causality graphs, with each subgraph having a tree-like structure and being rooted in the application frontend. 
The idea is that an anomaly originates in some application services and propagates from such services to the application frontend. 
All identified subgraphs are assigned a score based on the frequency with which they appear in the causality graphs. 
Only the subgraphs whose score is higher than a given threshold are kept, as they are considered as possible explanations for the performance anomaly observed on the application frontend.
FacGraph \cite{44_Lin2018_FacGraph} then returns the set of services corresponding to leaves in the tree-like structure of each of the kept subgraphs, which constitute the possible root causes for the observed frontend anomaly.

{%
Differently from FacGraph \cite{44_Lin2018_FacGraph}, LOUD \cite{13_Mariani2018_LocalizingFaultsCloudSystems} determines the root causes of service-level performance anomalies by building causality graphs whose vertices correspond to KPIs monitored on application services (rather than to the services themselves).
In addition, LOUD \cite{13_Mariani2018_LocalizingFaultsCloudSystems} is \enquote{KPI-agnostic}, as it works with any set of KPIs for monitored on the services forming an application.
LOUD \cite{13_Mariani2018_LocalizingFaultsCloudSystems} relies on the IBM ITOA-PI \cite{IBM_ITOA_PI} to detect anomalous KPIs monitored on the services forming an application (\Cref{sec:detection:monitoring:unsupervised-learning}), and to automatically determine the possible root causes for such anomalies.
The IBM ITOA-PI \cite{IBM_ITOA_PI} is indeed exploited to automatically build a causality graph and to process such graph.
This is done under the assumption that the anomalous KPIs related to a detected anomaly are highly correlated and form a connected subgraph of the causality graph.
In particular, LOUD assumes that the anomalous behaviour of the service whose KPIs are anomalous is likely to result in the anomalous behaviour of services it interacts with, either directly or through some other intermediate services.
Based on this assumption, LOUD exploits graph centrality indices to identify the anomalous KPIs that best characterize the root cause of a detected performance anomaly: the KPIs with the highest centrality scores likely correspond to the services that are responsible for the detected anomaly.
}

\subsection{Discussion}
\label{sec:rca:discussion}
\Cref{tab:rca} recaps the surveyed root cause analysis techniques by distinguishing their classes, \viz whether they are log-based, distributed tracing-based, or monitoring-based, as well as the method they apply to determine the possible root causes for an observed anomaly.
The table classifies the surveyed techniques based on whether they identify the possible root causes for functional or performance anomalies, observed at application-level (\eg on the application frontend) or on specific services in the application, as well as on whether they are already integrated with any of the anomaly detection techniques surveyed in \Cref{sec:detection}.
\Cref{tab:rca} also provides some insights on the artifacts that must be provided to the surveyed root cause analysis techniques, as input needed to actually search for possible root causes of observed anomalies, and it recaps the root causes they automatically identify.
In the latter perspective, the table distinguishes whether the surveyed techniques identify the services, events, or KPIs that possibly caused an observed anomaly, and whether the identified root causes are ranked, with higher ranks assigned to those that have higher probability to have caused the observed anomaly.

Taking \Cref{tab:rca} as a reference, we hereafter summarise the surveyed techniques for identifying the possible root causes for detected anomalies (\Cref{sec:rca:discussion:summary}).
We also discuss them under three different perspectives, \viz finding a suitable compromise between the identified root causes and setup costs (\Cref{sec:rca:discussion:root-causes-costs}), their accuracy (\Cref{sec:rca:discussion:false-positives-negatives}), and the need for explainability and countermeasures (\Cref{sec:rca:discussion:explainability-countermeasures}). 

\subsubsection{Summary}
\label{sec:rca:discussion:summary}
The surveyed root cause analysis techniques allow to determine the possible root causes for anomalies observed on multi-service applications.
In some cases, the root cause analysis techniques are enacted in pipeline with anomaly detection.
Whenever an anomaly is detected to affect the whole application or one of its services, a root cause analysis is automatically started to determine the possible root causes for such an anomaly  \cite{16_Liu2020_TraceAnomaly,49_Shan2019_epsilonDiagnosis,20_Wang2020_RootCauseMetricLocation,04_Wu2020_MicroRCA,09_Wu2020_MicroserviceDiagnosisDeepLearning,14_Samir2019_DLA,40_Chen2020_CauseInfer,48_Chen2014_CauseInfer,06_Lin2018_Microscope,19_Guan2019_MicroscopeDemo,29_Wang2018_CloudRanger,13_Mariani2018_LocalizingFaultsCloudSystems}.
{%
In this case, the pipeline of techniques can be used as-is to automatically detect anomalies and their possible root causes.
}
The other techniques instead determine the possible root causes for anomalies observed with external monitoring tools, by end users, or by application operators \cite{10_Aggarwal2020_FaultsGoldenSignals,01_Zhou2021_FaultAnalysisDebuggingMicroservices,45_Guo2020_GMTA,41_Mi2013_PerformanceDiagnosisCloud,21_Kim2013_RootCauseDetectionSOA,47_Liu2021_MicroHECL,42_Nguyen2011_PAL,43_Nguyen2013_FChain,30_Thalheim2017_Sieve,11_Brandon2020_GraphBasedRootCauseAnalysis,07_Ma2019_MSRank,25_Ma2020_MSRankExtended,08_Ma2020_AutoMAP,44_Lin2018_FacGraph,05_Meng2020_RootCausesMicroservices,46_Qiu2020_CausalityMiningKnowledgeGraph}.
{%
By combining the information on the type and grain of detected/analysed anomalies available in \Cref{tab:detection,tab:rca}, application operators can determine which root cause analysis technique can be used in pipeline with a given anomaly detection technique, after applying the necessary integration to both techniques to let them successfully interoperate.
}

\begin{table}[tp]
\caption{Classification of root cause analysis techniques, based on their \textit{class} (\viz \textsf{L} for log-based techniques, \textsf{DT} for distributed tracing-based techniques, \textsf{M} for monitoring-based techniques), the applied \textit{method}, whether the \textit{detection} of analysed anomalies is integrated or done with external tools (\viz \textsf{I} for integrated pipelines, \textsf{E} for external tools), the \textit{type} (\viz \textsf{F} for functional anomalies, \textsf{P} for performance anomalies) and \textit{grain} of explained anomalies (\viz \textsf{A} for application-level anomalies, \textsf{S} for service-level anomalies), the identified \textit{root causes}, and the \textit{input} they need to run.}
\label{tab:rca} 

\begin{tiny}
\begin{tabular}{%
    >{\centering\arraybackslash}m{.13\textwidth}%
    @{}
    >{\centering\arraybackslash}m{.05\textwidth}%
    @{\ }
    >{\centering\arraybackslash}m{.23\textwidth}%
    >{\centering\arraybackslash}m{.04\textwidth}%
    @{\,}
    >{\centering\arraybackslash}m{.04\textwidth}%
    @{\,}
    >{\centering\arraybackslash}m{.04\textwidth}%
    @{\,}
    >{\centering\arraybackslash}m{.14\textwidth}%
    @{\,}
    >{\centering\arraybackslash}m{.25\textwidth}%
}
    \thickline
        &
        & 
        & \multicolumn{3}{c}{\textbf{Anomaly}}
        &
        \\
        & \textbf{Class} 
         & \textbf{Method}
        & \textbf{Det.}
        & \textbf{Type}
        & \textbf{Gran.}
        & \textbf{Root Causes}
        & \textbf{Needed Input}
        \\ 
    \thickline
    Aggarwal \etal \cite{10_Aggarwal2020_FaultsGoldenSignals}
        & \textsf{L}
        & \cellcolor[HTML]{DFEEF6} random walk on causality graphs
        & \textsf{E}
        & \textsf{F}
        & \textsf{A}
        & service
        & runtime logs, app spec
        \\
    \thinline
    Zhou \etal \cite{01_Zhou2021_FaultAnalysisDebuggingMicroservices}
        & \textsf{DT}
        & \cellcolor[HTML]{FFD1C2} visual trace comparison
        & \textsf{E}
        & \textsf{F}, \textsf{P}
        & \textsf{A}
        & events 
        & previous and runtime traces
        \\
    GMTA \cite{45_Guo2020_GMTA}
        & \textsf{DT}
        & \cellcolor[HTML]{FFD1C2} visual trace comparison
        & \textsf{E}
        & \textsf{F}, \textsf{P}
        & \textsf{A}
        & services
        & previous and runtime traces
        \\
    
    CloudDiag \cite{41_Mi2013_PerformanceDiagnosisCloud}
        & \textsf{DT}
        & \cellcolor[HTML]{F3EDCE} RPCA on traces
        & \textsf{E}
        & \textsf{P}
        & \textsf{A}
        & ranked services
        & app deployment
        \\
    
    TraceAnomaly \cite{16_Liu2020_TraceAnomaly}
        & \textsf{DT}
        & \cellcolor[HTML]{E6D6C7} direct KPI correlation
        & \textsf{I}
        & \textsf{F}, \textsf{P}
        & \textsf{A}
        & services
        & previous and runtime traces
        \\
    MonitorRank \cite{21_Kim2013_RootCauseDetectionSOA}
        & \textsf{DT}
        & \cellcolor[HTML]{D2EEE3} random walk on topology graphs
        & \textsf{E}
        & \textsf{P}
        & \textsf{A}
        & ranked services
        &   app deployment
        \\
    MicroHECL \cite{47_Liu2021_MicroHECL}
        & \textsf{DT} 
        & \cellcolor[HTML]{C4E9DA} BFS on topology graphs
        & \textsf{E}
        & \textsf{P}
        & \textsf{S}
        & ranked services
        & previous and runtime traces 
        \\
    \thinline
    $\epsilon$-diagnosis \cite{49_Shan2019_epsilonDiagnosis}
        & \textsf{M}
        & \cellcolor[HTML]{E6D6C7} direct KPI correlation
        & \textsf{I}
        & \textsf{P}
        & \textsf{A}
        & service KPIs
        & k8s deployment
        \\
    
    Wang \etal \cite{20_Wang2020_RootCauseMetricLocation}
        & \textsf{M}
        & \cellcolor[HTML]{E6D6C7}  direct KPI correlation
        & \textsf{I}
        & \textsf{F}, \textsf{P}
        & \textsf{S}
        & service KPIs
        & runtime logs, monitored KPIs
        \\
    PAL \cite{42_Nguyen2011_PAL}
        & \textsf{M}
        & \cellcolor[HTML]{E6D6C7}  direct KPI correlation
        & \textsf{E}
        & \textsf{P}
        & \textsf{A}
        & ranked services
        & app deployment 
        \\
    FChain \cite{43_Nguyen2013_FChain}
        & \textsf{M}
        & \cellcolor[HTML]{E6D6C7}  direct KPI correlation
        & \textsf{E}
        & \textsf{P}
        & \textsf{A}
        & ranked services
        & app deployment 
        \\
    MicroRCA \cite{04_Wu2020_MicroRCA}
        & \textsf{M}
        & \cellcolor[HTML]{D2EEE3} random walk on topology graphs
        & \textsf{I}
        & \textsf{P}
        & \textsf{S}
        & services
        & k8s deployment, workload generator
        \\
    Wu \etal \cite{09_Wu2020_MicroserviceDiagnosisDeepLearning}
        & \textsf{M}
        & \cellcolor[HTML]{D2EEE3} random walk on topology graphs
        & \textsf{I}
        & \textsf{P}
        & \textsf{S}
        & service KPIs
        & k8s deployment, workload generator
        \\
    Sieve \cite{30_Thalheim2017_Sieve}
        & \textsf{M}
        & \cellcolor[HTML]{B5E3D1} KPI correlation on topology graphs
        & \textsf{E}
        & \textsf{P}
        & \textsf{A}
        & service KPIs
        & app deployment
        \\
    Brand\'on \etal \cite{11_Brandon2020_GraphBasedRootCauseAnalysis}
        & \textsf{M}
        & \cellcolor[HTML]{A6DDC8} pattern matching on topology graphs
        & \textsf{E}
        & \textsf{F}, \textsf{P}
        & \textsf{S}
        & ranked services
        & app deployment, anomaly patterns
        \\
    DLA \cite{14_Samir2019_DLA}
        & \textsf{M}
        & \cellcolor[HTML]{97D8BF} Markov analysis on topology graphs
        & \textsf{I}
        & \textsf{P}
        & \textsf{S}
        & service KPIs
        & k8s deployment, app spec
        \\
    CauseInfer \cite{40_Chen2020_CauseInfer,48_Chen2014_CauseInfer}
        & \textsf{M}
        & \cellcolor[HTML]{E7F2F8} BFS on causality graphs
        & \textsf{I}
        & \textsf{P}
        & \textsf{A}
        & ranked services
        & target  app deployment
        \\
    Microscope \cite{06_Lin2018_Microscope,19_Guan2019_MicroscopeDemo}
        & \textsf{M}
        & \cellcolor[HTML]{E7F2F8} BFS on causality graphs
        & \textsf{I}
        & \textsf{P}
        & \textsf{A}
        & ranked services
        & target  app deployment
        \\
    Qiu \etal \cite{46_Qiu2020_CausalityMiningKnowledgeGraph}
        & \textsf{M}
        & \cellcolor[HTML]{E7F2F8} BFS on causality graphs
        & \textsf{E}
        & \textsf{P}
        & \textsf{S}
        & ranked service KPIs
        & monitored KPIs, app spec
        \\
    CloudRanger \cite{29_Wang2018_CloudRanger}
        & \textsf{M}
        & \cellcolor[HTML]{DFEEF6} random walk on causality graphs
        & \textsf{I}
        & \textsf{P}
        & \textsf{A}
        & ranked services
        & app deployment, prev. monitored KPIs
        \\
    MS-Rank \cite{07_Ma2019_MSRank,25_Ma2020_MSRankExtended}
        & \textsf{M}
        & \cellcolor[HTML]{DFEEF6} random walk on causality graphs
        & \textsf{E}
        & \textsf{P}
        & \textsf{A}
        & ranked services
        & monitored KPIs
        \\
    AutoMAP \cite{08_Ma2020_AutoMAP}
        & \textsf{M}
        & \cellcolor[HTML]{DFEEF6} random walk on causality graphs
        & \textsf{E}
        & \textsf{P}
        & \textsf{A}
        & ranked services
        & monitored KPIs
        \\
    MicroCause \cite{05_Meng2020_RootCausesMicroservices}
        & \textsf{M}
        & \cellcolor[HTML]{DFEEF6} random walk on causality graphs
        & \textsf{E}
        & \textsf{P}
        & \textsf{S}
        & ranked service KPIs
        & monitored KPIs
        \\
    FacGraph \cite{44_Lin2018_FacGraph}
        & \textsf{M}
        & \cellcolor[HTML]{D0E5F1} correlation of causality graphs
        & \textsf{E}
        & \textsf{P}
        & \textsf{A}
        & ranked services
        & monitored KPIs
        \\
    LOUD \cite{13_Mariani2018_LocalizingFaultsCloudSystems}
        & \textsf{M}
        & \cellcolor[HTML]{C0DDED} centrality on causality graphs
        & \textsf{I}
        & \textsf{P}
        & \textsf{S}
        & ranked service KPIs
        & app deployment, workload generator
        \\
    \thickline
\end{tabular}
\end{tiny}
\end{table}

Whichever is the way of detecting anomalies, their root cause analysis is typically enacted on a graph modelling the multi-service application where anomalies have been observed (\Cref{tab:rca}).
A possibility is to visit a graph modelling the application topology, which can be automatically reconstructed by monitoring service interactions \cite{04_Wu2020_MicroRCA,09_Wu2020_MicroserviceDiagnosisDeepLearning,30_Thalheim2017_Sieve,11_Brandon2020_GraphBasedRootCauseAnalysis} or from distributed traces  \cite{21_Kim2013_RootCauseDetectionSOA,47_Liu2021_MicroHECL}, or which must be provided as input by the application operator \cite{14_Samir2019_DLA}.
The most common, graph-based approach is however reconstructing a causality graph from the application logs \cite{10_Aggarwal2020_FaultsGoldenSignals} or from the KPIs monitored on the services forming an application \cite{40_Chen2020_CauseInfer,48_Chen2014_CauseInfer,06_Lin2018_Microscope,19_Guan2019_MicroscopeDemo,29_Wang2018_CloudRanger,07_Ma2019_MSRank,25_Ma2020_MSRankExtended,08_Ma2020_AutoMAP,44_Lin2018_FacGraph,05_Meng2020_RootCausesMicroservices,46_Qiu2020_CausalityMiningKnowledgeGraph,13_Mariani2018_LocalizingFaultsCloudSystems}, typically to model how application services influence each other.
The vertices in a causality graph either model the application services \cite{40_Chen2020_CauseInfer,48_Chen2014_CauseInfer,06_Lin2018_Microscope,19_Guan2019_MicroscopeDemo,29_Wang2018_CloudRanger,07_Ma2019_MSRank,25_Ma2020_MSRankExtended,08_Ma2020_AutoMAP,44_Lin2018_FacGraph} or the KPIs \cite{05_Meng2020_RootCausesMicroservices,46_Qiu2020_CausalityMiningKnowledgeGraph,13_Mariani2018_LocalizingFaultsCloudSystems} monitored on such services, whereas each arc indicates that the performance of the target service/KPI depends on that of the source service/KPI.
Causality graphs are always obtained (but for the case of LOUD \cite{13_Mariani2018_LocalizingFaultsCloudSystems}) by applying the PC-algorithm \cite{Spirtes2000_CausationPredictionSearchPCAlgoritm} to the time series of logged events/monitored KPIs, as such algorithm is known to allow determining causal relationships in time series, \viz by identifying whether the values in a time series can be used to predict those of another time series.

The search can then be performed by visiting the topology/causality graph, starting from the application frontend in the case of application-level anomalies \cite{10_Aggarwal2020_FaultsGoldenSignals,06_Lin2018_Microscope,19_Guan2019_MicroscopeDemo,40_Chen2020_CauseInfer,48_Chen2014_CauseInfer,29_Wang2018_CloudRanger,07_Ma2019_MSRank,25_Ma2020_MSRankExtended,08_Ma2020_AutoMAP,44_Lin2018_FacGraph,21_Kim2013_RootCauseDetectionSOA,30_Thalheim2017_Sieve} or from the KPI/service where the anomaly was actually observed \cite{05_Meng2020_RootCausesMicroservices,46_Qiu2020_CausalityMiningKnowledgeGraph,13_Mariani2018_LocalizingFaultsCloudSystems,47_Liu2021_MicroHECL,04_Wu2020_MicroRCA,09_Wu2020_MicroserviceDiagnosisDeepLearning}.
Different methods are then applied to visit the graph, the most intuitive being a BFS over the graph to identify all possible paths that could justify the observed anomaly \cite{46_Qiu2020_CausalityMiningKnowledgeGraph}.
To reduce the paths to be visited, the BFS is often stopped when no anomaly was affecting any of the services that can be reached from the currently visited service, which is hence considered a possible root cause for the observed anomaly \cite{47_Liu2021_MicroHECL,40_Chen2020_CauseInfer,48_Chen2014_CauseInfer,06_Lin2018_Microscope,19_Guan2019_MicroscopeDemo}.
The most common method is however that first proposed in MonitorRank \cite{21_Kim2013_RootCauseDetectionSOA} and then reused by various techniques \cite{04_Wu2020_MicroRCA,09_Wu2020_MicroserviceDiagnosisDeepLearning,29_Wang2018_CloudRanger,07_Ma2019_MSRank,25_Ma2020_MSRankExtended,08_Ma2020_AutoMAP,05_Meng2020_RootCausesMicroservices}.
In this case, the graph is visited through a random walk, where the probability of visiting a service with a random step is proportional to the correlation between the performances of such service and of the service where the anomaly was observed. 
Rather than returning all possible root causes (as per what modelled by the graph), random walk-based techniques return a ranked list of possible root causes, under the assumption that the more a service gets visited through the random walk, the more is probable that it can have caused the observed anomaly \cite{21_Kim2013_RootCauseDetectionSOA,07_Ma2019_MSRank,25_Ma2020_MSRankExtended}.

Other graph-based root cause analysis techniques anyhow exist, either using the application topology to drive the pairwise comparison of monitored performances \cite{30_Thalheim2017_Sieve}, or based on processing the whole topology/causality graph to determine the possible root causes for an observed anomaly
\cite{14_Samir2019_DLA,11_Brandon2020_GraphBasedRootCauseAnalysis,13_Mariani2018_LocalizingFaultsCloudSystems}.
As for processing the application topology, the alternatives are twofold, \viz enacting pattern matching to reconduct the observed anomaly to some already troubleshooted situation \cite{11_Brandon2020_GraphBasedRootCauseAnalysis}, or transforming the topology into a probabilistic model to elict the most probable root causes for the observed anomaly \cite{14_Samir2019_DLA}.
In the case of causality graph, instead, graph processing corresponds to computing graph centrality indexes to rank the KPIs monitored on services, assuming that highest ranked KPIs best characterise the observed anomaly \cite{13_Mariani2018_LocalizingFaultsCloudSystems}. 

Quite different methods are trace comparison and {direct KPI correlation}. 
In the case of trace comparison, the idea is to provide application operators with a graphical support to visualise and compare traces, in such a way that they can troubleshoot the traces corresponding to anomalous runs of an application \cite{01_Zhou2021_FaultAnalysisDebuggingMicroservices,45_Guo2020_GMTA}.
{Direct KPI correlation} is instead based the idea here is that an anomaly observed on a service can be determined by the co-occurrence of other anomalies on other services.
The proposed method is hence to {directly} process the traces produced by application services \cite{41_Mi2013_PerformanceDiagnosisCloud,16_Liu2020_TraceAnomaly} or their monitored KPIs \cite{49_Shan2019_epsilonDiagnosis,20_Wang2020_RootCauseMetricLocation,42_Nguyen2011_PAL,43_Nguyen2013_FChain} to detect co-occurring anomalies, \viz occurring in a time interval including the time when the initially considered anomaly was observed.
Such co-occurring anomalies are considered as the possible root causes for the observed anomaly.

\subsubsection{Identified Root Causes vs. Setup Costs}
\label{sec:rca:discussion:root-causes-costs}
The surveyed techniques determine possible root causes for anomalies of different types and granularity, by requiring different types of artifacts (\Cref{tab:detection}).
Independently from the applied method, trace-based and monitoring-based techniques reach a deeper level of detail for considered anomalies and identified root causes, if compared with the only available log-based technique.
The log-based technique by Aggarwal \etal \cite{10_Aggarwal2020_FaultsGoldenSignals} determines a root causing service, which most probably caused the functional anomaly observed on the frontend of an application.
The available trace-based and monitoring-based techniques instead determine multiple possible root causes for application-level anomalies \cite{01_Zhou2021_FaultAnalysisDebuggingMicroservices,45_Guo2020_GMTA,41_Mi2013_PerformanceDiagnosisCloud,16_Liu2020_TraceAnomaly,21_Kim2013_RootCauseDetectionSOA,49_Shan2019_epsilonDiagnosis,42_Nguyen2011_PAL,43_Nguyen2013_FChain,30_Thalheim2017_Sieve,40_Chen2020_CauseInfer,48_Chen2014_CauseInfer,06_Lin2018_Microscope,19_Guan2019_MicroscopeDemo,29_Wang2018_CloudRanger,07_Ma2019_MSRank,25_Ma2020_MSRankExtended,08_Ma2020_AutoMAP,44_Lin2018_FacGraph} or service-level anomalies \cite{47_Liu2021_MicroHECL,20_Wang2020_RootCauseMetricLocation,04_Wu2020_MicroRCA,09_Wu2020_MicroserviceDiagnosisDeepLearning,11_Brandon2020_GraphBasedRootCauseAnalysis,14_Samir2019_DLA,05_Meng2020_RootCausesMicroservices,46_Qiu2020_CausalityMiningKnowledgeGraph,13_Mariani2018_LocalizingFaultsCloudSystems}.
They can also provide more details on the possible root causes for an observed anomaly, from the application services that may have caused it \cite{45_Guo2020_GMTA,41_Mi2013_PerformanceDiagnosisCloud,16_Liu2020_TraceAnomaly,21_Kim2013_RootCauseDetectionSOA,47_Liu2021_MicroHECL,42_Nguyen2011_PAL,43_Nguyen2013_FChain,04_Wu2020_MicroRCA,11_Brandon2020_GraphBasedRootCauseAnalysis,40_Chen2020_CauseInfer,48_Chen2014_CauseInfer,29_Wang2018_CloudRanger,07_Ma2019_MSRank,25_Ma2020_MSRankExtended,08_Ma2020_AutoMAP,44_Lin2018_FacGraph} up to the actual events \cite{01_Zhou2021_FaultAnalysisDebuggingMicroservices} or KPIs monitored on such services \cite{49_Shan2019_epsilonDiagnosis,20_Wang2020_RootCauseMetricLocation,09_Wu2020_MicroserviceDiagnosisDeepLearning,30_Thalheim2017_Sieve,14_Samir2019_DLA,05_Meng2020_RootCausesMicroservices,46_Qiu2020_CausalityMiningKnowledgeGraph,13_Mariani2018_LocalizingFaultsCloudSystems}.
To help application operators in troubleshooting the multiple root causes identified for an observed anomaly, some techniques also rank the possible root causing services \cite{41_Mi2013_PerformanceDiagnosisCloud,21_Kim2013_RootCauseDetectionSOA,47_Liu2021_MicroHECL,42_Nguyen2011_PAL,43_Nguyen2013_FChain,11_Brandon2020_GraphBasedRootCauseAnalysis,40_Chen2020_CauseInfer,48_Chen2014_CauseInfer,06_Lin2018_Microscope,19_Guan2019_MicroscopeDemo,29_Wang2018_CloudRanger,07_Ma2019_MSRank,25_Ma2020_MSRankExtended,08_Ma2020_AutoMAP,44_Lin2018_FacGraph} or KPIs \cite{05_Meng2020_RootCausesMicroservices,46_Qiu2020_CausalityMiningKnowledgeGraph,13_Mariani2018_LocalizingFaultsCloudSystems}.
The idea is that highest ranked root causes are more likely to have caused an observed anomaly, hence helping application operators to earlier identify the actual root causes for the observed anomaly, often saving them from troubleshooting all possible root causes.

At the same time, the log-based root cause analysis proposed by Aggarwal \etal \cite{10_Aggarwal2020_FaultsGoldenSignals} directly works with the event logs produced by application services, by inferring causal relationships among logged events based on an application specification provided by the application operator. 
The log-based root cause analysis technique proposed by Aggarwal \etal \cite{10_Aggarwal2020_FaultsGoldenSignals} hence assumes that application services log events, which is typically the case, whilst not requiring any further application instrumentation.
As we already discussed in \Cref{sec:detection:discussion:type-granularity-costs}, distributed tracing-based and monitoring-based techniques instead require applications to be instrumented to feature distributed tracing or to be deployed together with agents monitoring KPIs on their services, respectively.
The surveyed techniques also often rely on specific technologies, such as k8s \cite{49_Shan2019_epsilonDiagnosis,04_Wu2020_MicroRCA,09_Wu2020_MicroserviceDiagnosisDeepLearning,14_Samir2019_DLA}, assume the deployment of an application to be such that each service is deployed on a different VM \cite{42_Nguyen2011_PAL,43_Nguyen2013_FChain}, or require the application operator to provide additional artifacts, \eg a specification of the application topology \cite{10_Aggarwal2020_FaultsGoldenSignals,14_Samir2019_DLA,46_Qiu2020_CausalityMiningKnowledgeGraph}, known anomaly patterns \cite{11_Brandon2020_GraphBasedRootCauseAnalysis}, or workload generators \cite{04_Wu2020_MicroRCA,09_Wu2020_MicroserviceDiagnosisDeepLearning,13_Mariani2018_LocalizingFaultsCloudSystems}.

The surveyed root cause analysis techniques hence differ in the type and granularity of observed anomalies they can explain, in the root causes they identify, and in their applicability to a multi-service application as it is.
In particular, whilst some techniques can achieve a deeper level of detail in identified root causes and rank them based on their probability to have caused an observed anomaly, this is paid with the cost for setting them up to work with a given application, \viz to suitably instrument the application or its deployment, or to provide the needed artifacts.
The choice of which technique to use for enacting root cause analysis on a given application hence consists of finding a suitable trade-off between the level and ranking of identified root causes and the cost for setting up such technique to work with the given application. 
We believe that the classification provided in this paper can provide a first support for application operators wishing to choose the root cause analysis technique most suited to their needs.

\subsubsection{Accuracy of Enacted Root Cause Analyses}
\label{sec:rca:discussion:accuracy}
\label{sec:rca:discussion:false-positives-negatives}
As for the case of anomaly detection (\Cref{sec:detection:discussion:adaptability}), false positives and false negatives can affect the accuracy of root cause analysis techniques.
False positives here correspond to the services or KPIs that are identified as possible root causes for an observed anomaly, even if they actually have nothing in common with such anomaly.
The only situation in which no false positive is returned by a root cause analysis technique is when Aggarwal \etal \cite{10_Aggarwal2020_FaultsGoldenSignals} returns the service that is actually causing an observed anomaly.
In all other cases, false positives are there: if Aggarwal \etal \cite{10_Aggarwal2020_FaultsGoldenSignals} returns a service different from that actually causing an observed anomaly, the returned service is a false positive.
The other surveyed techniques instead return multiple possible root causes, often including various false positives.
Such false positives put more effort on the application operator, who is required to troubleshoot the corresponding services, even if they actually not caused the anomaly whose root causes are being searched.

At the same time, returning multiple possible root causes reduces the risk for false negatives, \viz for the actual root causes of an observed anomaly to not be amongst those determined by the enacted root cause analysis.
The impact of false negatives is even more severe than that of false positives, which puts more effort on the application operator \cite{46_Qiu2020_CausalityMiningKnowledgeGraph}.
\upd{The main purpose of root cause analysis is indeed to automatically identify the actual root causes for an observed anomaly, and missing such actual root causes (as in the case of false negatives) may make the price to enact root cause analysis not worthy to be paid.}
This is the main reason why most of the surveyed techniques opt to return multiple possible root causes, amongst which the actual one is likely to be.
As we already highlighted in \Cref{sec:rca:discussion:summary}, to reduce the effort to be put on the application operator to identify the actual root cause among the returned ones, some techniques also rank them based on their likelihood to have caused the observed anomaly \cite{21_Kim2013_RootCauseDetectionSOA}.

False negatives are however always around the corner.
The common driver for automatically identifying root causes with the surveyed techniques is correlation, which is used to correlate offline detected anomalies or to drive the search within the application topology or in a causality graph.
It is however known that correlation is however not ensuring causation \cite{Calude2017_SpuriousCorrelations}, hence meaning that the actual root cause for an anomaly observed in a service may not be amongst the services whose behaviour is highly correlated with that of the anomalous service, but rather in some other service.
Spurious correlations impact more on those techniques determining causalities by relying on correlation only, \eg to determine which is the root cause among co-occurring anomalies, or to build a causality graph.
One may hence think that topology-driven techniques are immune to spurious correlations, as they rely on the explicit modelling of the interconnections between the services forming an application and their hosting nodes. 
However, those techniques enacting correlation-guided random walks over the topology of an application are still prone to spurious correlations.
At the same time, even if a technique enacts an extensive search over the application topology and considers all modelled dependencies, it may miss the actual root cause in some cases.
For instance, if a topology is just representing service interactions, a technique analysing such topology may miss the case of performance anomalies affecting a service because other services hosted on the same node are consuming all computing resources.
It may also be the case that ---independently from the applied method--- a root cause technique actually determines the root cause for an observed anomaly, but the latter is excluded from those returned because its estimated likelihood to have caused the observed anomaly is below a given threshold.

All the surveyed techniques are hence subject to both false positives and false negatives, as they are actually inherent to the root cause analysis problem itself.
False positives/negatives are indeed a price to pay when enacting root cause analysis.
A quantitative comparison of the accuracy of the different techniques on given applications in given contexts would further help application operators in this direction, and it will complement the information in this survey in supporting application in choosing the root cause analysis techniques best suited to their needs.
Such a quantitative comparison is however outside of the scope of this survey and left for future work.

\subsubsection{Explainability and Countermeasures}
\label{sec:rca:discussion:explainability-countermeasures}
The problem of false positives/negatives amongst identified root causes can be mitigated by addressing one of the open challenges in root cause analysis: explainability.
Identified root causes should be associated with explanations on why they may have caused an observed anomaly, as well as why they are ranked higher than other possible root causes, in the case of techniques returning a ranking of possible root causes. 
Such information would indeed help application operators in identifying the false positives to get excluded, or in considering the possible root causes in some order different from that given by the ranking, if they believe that their associated explanations are more likely to have caused an observed anomaly.

In addition, as we discussed in \Cref{sec:rca:discussion:false-positives-negatives}, false negatives may be due to the fact that a root cause analysis technique excluded possible root causes if their likelihood to have caused an observed anomaly was below a given threshold.
If returned root causes would be associated with explanations on why they were considered such, we may avoid cutting the returned set of possible root causes, hence reducing the risk of techniques to consider as false negatives those root causes that they actually determined.
Again, application operators could exploit provided explanations to directly exclude the root causes corresponding to false positives.

Recommending countermeasures to be enacted to avoid the identified root causes to cause again the correspondingly observed anomalies is also an open challenge.
All the surveyed techniques allow to determine possible root causes for an observed anomaly, \viz the set of services that may have caused the anomaly, or the trace events or KPIs collected on such services. 
The possible root causes are returned to the application operator, who is in charge of further troubleshooting the potentially culprit services to determine whether/why they actually caused the observed anomaly.
At the same time, the possible root causes are identified based on data collected on the services forming an application, which is processed only to enact root cause analysis.
Such data could however be further processed to further support the application operator, \eg to automatically extract possible countermeasures to avoid it to cause the observed anomaly \cite{Brogi2020_FailureCausalities}.
For instance, if a service logged an error event that caused a cascade of errors reaching the service where the anomaly was observed, the root cause analysis technique could further process the error logs to suggest how to avoid such error propagation to happen again, \eg by introducing circuit breakers or bulkheads \cite{Richardson2018_MicroservicesPatterns}.
If a service was responding too slowly and caused some performance anomaly in another service, this information can be used to train machine learning models to predict similar performance degradations in the root causing service, which could then be used to preemptively scale such service and avoid the performance degradation to again propagate in future runs of an application.
The above are just two out of many possible examples, deserving further investigation and opening new research directions. 

\section{Related Work}
\label{sec:related}
Anomaly detection and root cause analysis are crucial in nowadays enterprise IT applications \cite{16_Liu2020_TraceAnomaly,29_Wang2018_CloudRanger}.
Techniques for solving both problems have already been discussed in existing surveys, which however differ from ours as they consider either anomaly detection or root cause analysis.

Chandola \etal \cite{Chandola2009_SurveyAnomalyDetection} and Akoglu \etal \cite{Akoglu2015_GraphBasedAnomalyDetection} survey the research on detecting anomalies in data.
\upd{In both surveys, the focus is on the generic problem of significant changes/outlier detection in data, independently of whether this is done online or offline, or what the data is about.
We differ from both surveys as we focus on online anomaly detection in a specific type of data, \viz logs, traces, or KPIs monitored on application services.
This allows us to provide a more detailed description of how such techniques can detect anomalies in running multi-service applications, as well as to compare and discuss them under dimensions peculiar to such problem, \eg whether they can detect application- or service-level anomalies, the cost for their setup, or their accuracy in dynamically changing applications.}
In addition, the studies surveyed by Chandola \etal \cite{Chandola2009_SurveyAnomalyDetection} and Akoglu \etal \cite{Akoglu2015_GraphBasedAnomalyDetection} are dated 2015 at most, hence meaning that various anomaly detection techniques for multi-service applications are not covered by such surveys.
Such techniques have indeed been proposed after 2014, when microservices were first proposed \cite{Lewis2014_Microservices} and gave rise to the widespread use of multi-service architectures \cite{Soldani2018_MicroservicesPainsGains}.
Finally, differently from Chandola \etal \cite{Chandola2009_SurveyAnomalyDetection} and Akoglu \etal \cite{Akoglu2015_GraphBasedAnomalyDetection}, we also survey the analysis techniques that can be used in a pipeline with the surveyed anomaly detection techniques to determine the possible root causes for detected anomalies.

Steinder and Sethi \cite{Steinder2004_FaultLocalizationTechniquesComputerNetworks}, Wong \etal \cite{Wong2016_SurveySoftwareFaultLocalization}, and Sole \etal \cite{Sole2017_SurveyRootCauseAnalysis} survey solutions for determining the possible root causes for failures in software systems.
In particular, Steinder and Sethi \cite{Steinder2004_FaultLocalizationTechniquesComputerNetworks} survey techniques for determining the root causes of anomalies in computer networks.
Wong \etal \cite{Wong2016_SurveySoftwareFaultLocalization} instead provide an overview of the techniques allowing to identify the root causes of failures in the source code of a single software program.  
\upd{Both Steinder and Sethi \cite{Steinder2004_FaultLocalizationTechniquesComputerNetworks} and Wong \etal \cite{Wong2016_SurveySoftwareFaultLocalization} hence
differ from our survey as they focus on software systems different than multi-service applications}.
In this perspective, the survey by Sole \etal \cite{Sole2017_SurveyRootCauseAnalysis} is closer to ours, since they focus on techniques allowing to determine the root causes of anomalies observed in multi-component software systems, therein included multi-service applications.
The focus of Sole \etal \cite{Sole2017_SurveyRootCauseAnalysis} is however complementary to ours: they analyse the performance and scalability of the surveyed root cause analysis techniques, whilst we focus on more qualitative aspects, \eg what kind of anomalies can be analysed, which analysis methods are enacted, which instrumentation they require, or which actual root causes are determined.
In addition, we also survey existing techniques for detecting anomalies in multi-service application, with the aim of supporting application operators in setting up their pipeline to detect anomalies in their applications and to determine the root causes of observed anomalies.

Another noteworthy secondary study is that by Arya \etal \cite{Arya2021_EvaluationCausalInferenceAIOps}, who evaluate several state-of-the-art techniques for root cause analysis, based on logs obtained from a publicly available benchmark multi-service applications. 
In particular, they consider Granger causality-based techniques, by providing a first quantitative evaluation of their performance and accuracy on a common dataset of application logs.
They hence complement the qualitative comparison in our survey with a first result along a research direction emerging from our survey, \viz quantitatively comparing the performance and accuracy of the existing root cause analysis techniques. 

In summary, to the best of our knowledge, ours is the first survey on the techniques for detecting anomalies that can be symptoms of failures in multi-service applications, while at the same time presenting the analysis techniques for determining possible root causes for such anomalies.
It is also the first survey providing a more detailed discussion of such techniques in the perspective of their use with multi-service applications, \eg whether they can detect and analyse anomalies at application- or service-level, which instrumentation they require, and how they behave when the services forming an application and their runtime environment change over time.

\section{Conclusions}
\label{sec:conclusions}
To support application operators in detecting failures and identiying their possible root causes, we surveyed existing techniques for anomaly detection and root cause analysis  in modern multi-service applications. 
Other than presenting the methods applied by such techniques, we discussed the type and granuality of anomalies they can detect/analyse, their setup costs (\viz the instrumentation that application operators must apply to their applications to use such techniques, and the additional artifacts they must produce), their accuracy and applicability to dynamically changing applications, and open challenges on the topic.

We believe that our survey can provide benefits to practicitioners working with modern multi-service applications, such as microservice-based applications, for instance. 
Our survey provides a first support for setting up a pipeline of anomaly detection and root cause analysis techniques, so that failures are automatically detected and analysed in multi-service applications.
We indeed not only highlighted which existing techniques already come in an integrated pipeline, but we also discussed the type and granularity of anomalies they consider. 
\upd{This provides a baseline for application operators to identify the anomaly detection and root cause analysis techniques that ---even if coming as independent solutions--- could be integrated in a pipeline, since they identify and explain the same type/granularity of anomalies}.
This, togeter with our discussion on the techniques' setup costs (\Cref{sec:detection:discussion:type-granularity-costs,sec:rca:discussion:root-causes-costs}), provides a first support to practitioners for choosing the most suited techniques for their needs.

In this perspective, a quantitative comparison of the performance of the surveyed techniques would further support practitioners, and it definitely deserves future work. 
Indeed, whilst the surveyed techniques already measure their performance or accuracy, sometimes also in comparison with other existing techniques, this is typically done with experiments on different reference applications running in different environments.
This makes it complex to quantitatively compare the existing anomaly detection and root cause analysis techniques, hence opening the need for studies quantitatively comparing the performance of such techniques by fixing the reference applications and runtime environments, much in a similar way as Arya \etal \cite{Arya2021_EvaluationCausalInferenceAIOps} did for a subset of existing root cause analysis techniques.  
Quantitative performance comparisons would complement the results of our qualitative comparison, providing practitioners with more tools to choose the anomaly detection and root cause analysis techniques most suited to their applications' requirements.

Our survey can also help researchers investigating failures and their handling in modern multi-service applications, as it provides a first structured discussion of anomaly detection and root cause analysis techniques in such a kind of applications.
For instance, we discussed how false positives and false negatives can negatively affect the accuracy of observed anomalies/identified root causes (\Cref{sec:detection:discussion:adaptability,sec:rca:discussion:accuracy}).
Whilst false positives could result in unnecessary work for application operators, false negatives could severely impact on the outcomes of the surveyed technique: false negatives indeed correspond to considering a service as not anomalous or not causing an observed anomaly, even if this was actually the case.
In both cases, false positives/negatives can occur if the services forming an application or its runtime conditions change over time.
Whilst some of the surveyed techniques already consider the issue of accuracy losses in the case of application changing over time, this is done with time consuming processes, \eg by re-training the machine learning models used to detect anomalies \cite{14_Samir2019_DLA,26_Wang2020_WorkflowAwareFaultDiagnosis,17_Gan2019_Seer} or by relying on application operators to provide updated artifacts to identify possible root causes \cite{11_Brandon2020_GraphBasedRootCauseAnalysis}.
Re-training or manual updates, as well as the other proposed updates, are hence not suited to get continuously enacted, but rather require to find a suitable tradeoff between the update period and the accuracy of enacted anomaly detection/root cause analysis \cite{14_Samir2019_DLA,17_Gan2019_Seer}.
However, modern software delivery practices are such that new versions of the services forming an application are continuously released \cite{Humble2010_ContinuousDelivery}, and cloud application deployments are such that services are often migrated from one runtime to another \cite{Carrasco2020_MigrationTransCloud}.  
An interesting research direction is hence to devise anomaly detection and root cause analysis techniques that can effectively work also in presence of continuous changes in multi-service applications, either adapting the surveyed ones or by devising new ones, \eg by exploiting the recently proposed continual learning solutions \cite{Lomonaco2021_ContinualAI}.

Other two interesting research directions already emerged in \Cref{sec:detection:discussion:explainability-countermeasures,sec:rca:discussion:explainability-countermeasures}, where we discussed the need for explainability and countermeasures.
On the one hand, associating observed anomalies and their identified root causes with explanations of why they are considered so would allow application operators to directly exclude false positives, which in turn means that, \eg we could avoid cutting the possible root causes for an observed anomaly to only the most probable ones.
Indeed, an application operator could be provided with all possible root causes and she could later decide which deserve to be troubleshooted based on their explanations.
This hence calls for anomaly detection and root cause analysis techniques that are \enquote{explainable by design}, much in the same way as the need for explainability is nowadays recognized in AI \cite{Guidotti2019_Explainability}.

On the other hand, recommending potential countermeasures to be enacted to avoid (the failures corresponding) to observed anomalies and their root causes would enable avoiding such failures to happen again in the future.
This could be done both for anomaly detection and for root cause analysis.
Indeed, whilst avoiding (the failure corresponding to a) detected anomaly of course needs acting on the root causes for such an anomaly, countermeasures could anyhow be taken to avoid the anomaly in a service to propagate to other services.
At the same time, the possible root causes for an observed anomaly are identified based on data collected on the services forming an application, which is processed only to enact root cause analysis.
Such data could however be further processed to automatically extract possible countermeasures to avoid it to cause the observed anomaly.
For instance, if a service logged an error event that caused a cascade of errors reaching the service where the anomaly was observed, the root cause analysis technique could further process the error logs to suggest how to avoid such error propagation to happen again, \eg by introducing circuit breakers or bulkheads \cite{Richardson2018_MicroservicesPatterns}.
If a service was responding too slowly and caused some performance anomaly in another service, this information can be used to train machine learning models to predict similar performance degradations in the root causing service, which could then be used to preemptively scale such service and avoid the performance degradation to again propagate in future runs of an application.
The above are just two out of many possible examples, deserving further investigation and opening a new research direction. 

\bibliographystyle{ACM-Reference-Format}
\bibliography{src/biblio}


\end{document}